\documentclass[twocolumn]{aastex631}

\usepackage{newtxtext,newtxmath}
\usepackage[T1]{fontenc}
\DeclareRobustCommand{\VAN}[3]{#2}
\let\VANthebibliography\thebibliography
\def\thebibliography{\DeclareRobustCommand{\VAN}[3]{##3}\VANthebibliography}
\usepackage{graphicx}	
\usepackage{amsmath}	

\usepackage{amssymb}	
\usepackage{mathtools}

\begin{document}

\title[Thirty Two  New ZZ Cetis from {\em TESS}]{Thirty Two New Bright ZZ Ceti Stars from {\em TESS}: Adding Cycles 4 and 5 }

\author[0000-0002-0797-0507]{Alejandra D. Romero}
\affiliation{Instituto de F\'{\i}sica, Universidade Federal do Rio Grande do Sul, 91501-970 Porto Alegre, RS, Brazil}

\author{S. O. Kepler}
\affiliation{Instituto de F\'{\i}sica, Universidade Federal do Rio Grande do Sul, 91501-970 Porto Alegre, RS, Brazil}

\author{Gabriela Oliveira da Rosa}
\affiliation{Instituto de F\'{\i}sica, Universidade Federal do Rio Grande do Sul, 91501-970 Porto Alegre, RS, Brazil}

\author{J. J. Hermes}
\affiliation{Department of Astronomy \& Institute for Astrophysical Research, Boston University, 725 Commonwealth Ave., Boston, MA 02215, USA}

\begin{abstract}

Analyzing all 120~s and 20~s light curves obtained by the TESS satellite up to Sector 69 --- the end of the fifth year of observations --- for all known white dwarfs and white dwarf candidates up to G=17.5 mag, we report the discovery of 32 new pulsating DA white dwarf stars. For all objects, we obtained the period spectra and performed a seismological analysis using full evolutionary models to estimate the structural parameters, such as effective temperature, stellar mass, and hydrogen envelope mass. The median stellar mass for the pulsators from asteroseismology is 0.609~M$_\odot$, in agreement with the median value from photometric and spectroscopic determinations, 0.602~M$_\odot$, excluding the low and extremely-low mass objects. Finally, we found rotational-splitting multiplets for 9 stars, which led to rotation periods between 4~h and 1~d.

\end{abstract}


\section{Introduction} \label{sec:intro}

The first variable white dwarf star was discovered by Arlo Landolt around 1960, when he was conducting photometric observations of faint standard stars at the Kitt Peak National Observatory. Landolt realized that one of the comparison stars, Haro--Luyten Taurus 76 or HL Tau 76, was actually variable itself, with a period of $\sim$750~s \citep{1968ApJ...153..151L}. The second object was reported in 1971 by \citet{1971ApJ...163L..89L}, who discovered photometric variability in the known DA white dwarf stars R548. Curiously, this object was already identified as a variable star and had been given the variable star name ZZ Ceti. R584 then became the prototype of the new class of pulsating white dwarfs known as ZZ Cetis \citep{2015uswd.book.....V}. By 2007, around 126 ZZ Cetis had been discovered \citep{2007CoAst.150..221K}. To date, approximately 500 ZZ Cetis have been confirmed \citep[see for instance][]{2016IBVS.6184....1B,2019A&ARv..27....7C, 2019MNRAS.490.1803R, 2020AJ....160..252V,2021ApJ...912..125G, Romero22}.

Variable DA white dwarfs are found in a narrow range of effective temperature, between 10,000~K and 13,000~K, depending on stellar mass \citep{2017ApJS..232...23H,2017EPJWC.15201011K}. Pulsations are identified as nonradial gravity spheroidal modes with low harmonic degrees, with variation periods ranging from 70 to 2000 seconds and amplitudes up to 0.3 magnitudes \citep{2016IBVS.6184....1B}. Pulsations are driven by an excitation mechanism known as the $\kappa$-$\gamma$ mechanism, related to an opacity bump due to partial hydrogen ionization at the base of the hydrogen-rich envelope \citep{1981A&A...102..375D,1982ApJ...252L..65W}. As the star cools, the outer convection layer thickens and the convective driving mechanism begins to dominate \citep{1991MNRAS.251..673B,1999ApJ...511..904G}.

ZZ Ceti stars can be classified into three categories: hot, intermediate, and cool, based on their effective temperature and pulsational properties \citep{1993BaltA...2..407C,2006ApJ...640..956M}.
Hot ZZ Cetis are found at the blue edge of the instability strip and exhibit stable light curves with short periods (less than 350 seconds) and small amplitudes (1.5 to 20 millimagnitudes). Cool ZZ Cetis, on the other hand, are situated at the red edge of the instability strip. They show long periods (up to 1500 s) with larger variation amplitudes (40 to 110 millimagnitudes) and nonsinusoidal light curves characterized by mode interference. Intermediate ZZ Cetis share the characteristics of both hot and cold members. Usually, for periods longer than 800 s, numerous peaks are detected in the Fourier transform, under a broad envelope that can be related to stochastically driven oscillations \citep{2020ApJ...890...11M}. Finally, the number of detected periods seems to increase at lower effective temperatures. 

The Transiting Exoplanet Survey Satellite (TESS) was launched on April 18, 2018, its primary mission being to search for exoplanets around bright stars with the transit method \citep{2014SPIE.9143E..20R}. During the first five cycles, the TESS mission has contributed significantly to the study of stellar pulsations in compact objects, in particular for hydrogen-rich atmosphere DA white dwarfs, through continuous and stable photometry, as well as its wide sky coverage \citep{2020A&A...638A..82B, 2020ApJ...888...49W, Romero22, 2023A&A...674A.204B}. During the first two cycles, 120-second cadence observations were available for selected targets, while for the extended mission for 2020-2022 many objects were observed at 20-second cadence, providing the opportunity to resolve the shortest pulsation periods of hot ZZ Ceti stars.  The activities related to compact pulsators are coordinated by the Evolved Compact Stars with TESS Working Group (WG8) of the TESS Asteroseismic Science Consortium.

\citet{Romero22} presented the discovery of 74 new ZZ Cetis based on TESS data from the first three cycles, from Sectors 1 to 39, including data from 120\,s and 20\,s cadence. They complemented the space-based data with ground-based observations for 11 objects, leading to the discovery of an additional ZZ Ceti, TIC 20979953, which was part of the target list for the TESS mission, but no data had been taken. \citet{Romero22} also detected multiplets for four objects in the sample, leading to rotation periods from one to a few hours. In this paper, we continue their work by including in our analysis the data taken by the TESS mission during its fourth and fifth years, from Sectors 40 to 69 with 120\,s- and 20\,s-cadence. As a result, we present the discovery of 32 new ZZ Cetis. In addition, we present an updated period list for 21 objects reported by \citet{Romero22}. 

This paper is organized as follows. We present the sample of the 53 ZZ Cetis analyzed in this work, including the 32 new pulsating DA white dwarfs, in Section~\ref{section2}. We describe sample selection and data reduction for the TESS data in Section~\ref{section3}. In Section~\ref{section4} we present the detected pulsation periods and perform an asteroseismological study for our sample in Section~\ref{section5}. In Section~\ref{section6} we present a study of the sample of 103 pulsating DA white dwarfs discovered using data from the first five years of the {\it TESS} mission. We conclude in Section~\ref{conclusions} by summarizing our findings.

\section{New ZZ Ceti stars}
\label{section2}

As a continuation of the work of \citet{Romero22}, we analyse the data from the fourth and fifth cycles of TESS, corresponding to Sectors 40 to 69, including data from 120\,s and 20\,s cadence. We look for newly observed ZZ Ceti stars, as well as for new data for objects that were already visited in previous cycles. As a result, we report 32 new ZZ Ceti stars, 12 new ZZ Cetis confirmed as pulsators with newly added data, and 20 objects observed only during the fourth and fifth cycles. Finally, 21 ZZ Cetis reported as variables in \citet{Romero22} were also observed in Sectors 40 to 69, and for those we update their period list. All targets analyzed are listed in Table~\ref{tab:list1}, where we include the coordinates in J2000, the $G$ magnitude, the effective temperature, surface gravity, and stellar mass. The last column indicates whether the object was reported as a ZZ Ceti by \citet{Romero22} (R22), or was confirmed as a pulsating DA white dwarf after observations from Sectors 40 to 69 were obtained (new).

The values for surface gravity and effective temperature listed in Table \ref{tab:list1} were taken from various works, which used different techniques to determine atmospheric parameters (see the references in column 8). 
Most of the data is taken from \citet{2019MNRAS.482.4570G,2021MNRAS.508.3877G}, and were determined using the magnitudes of {\it Gaia} DR2 and DR3 and parallax, combined with hydrogen models from \citet{2011ApJ...730..128T}. 

Photometric determinations of atmospheric parameters were also taken from the works of \citet{2020ApJ...898...84K} and \citet{2020AJ....160..252V}. The authors used parallaxes from {\it Gaia} DR2 and photometry from the Sloan Digital Sky Survey \citep{2006ApJS..167...40E,2013ApJS..204....5K,2019MNRAS.486.2169K} and the Panoramic Survey Telescope and Rapid Response System \citep{2016arXiv161205560C}. They used pure hydrogen atmosphere models from \citet{2019ApJ...876...67B}, applying the photometric technique described in \citet{1997ApJS..108..339B}. Spectroscopic determinations of the effective temperature and surface gravity were taken from \citet{2011ApJ...743..138G,2013AJ....145..136L,2017MNRAS.472.4173R, 2019MNRAS.486.2169K} and \citet{2020MNRAS.497..130T}. The stellar mass values were estimated by linear interpolation of evolutionary tracks in the cooling sequences described in \citet{2019MNRAS.490.1803R} in the $\log g$ – $T_{\rm  eff}$ plane, given the values of surface gravity and effective
temperature from Table~\ref{tab:list1}. 

\begin{figure*}
	\includegraphics[width=0.9\textwidth]{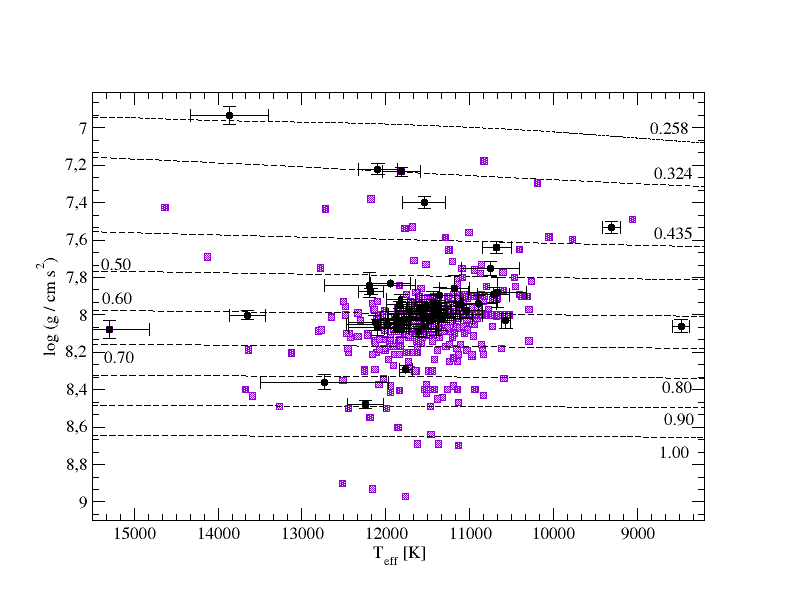}
    \caption{Distribution of ZZ Ceti stars on the $T_{\rm  eff}-\log g$ plane. The purple squares correspond to the ZZ Ceti stars taken from \citet{2016IBVS.6184....1B, 2017ApJS..232...23H, 2017ApJ...847...34S, 2019MNRAS.490.1803R, 2020AJ....160..252V} and \citet{Romero22}. The objects of interest in this work are depicted with black circles. We include evolutionary tracks (dashed lines) with stellar masses between 0.435~M$_{\odot}$ and 0.9~M$_{\odot}$ from \citet{2019MNRAS.484.2711R} and 0.258 and 0.324~M$_{\odot}$ from \citet{2014A&A...571A..45I}.} 
    \label{ZZCetis}
\end{figure*}

We show the location of all the objects listed in Table \ref{tab:list1} in the T$_{\rm eff}-\log g$ plane in Figure \ref{ZZCetis} (black circles). For targets that have more than one determination of atmospheric parameters, we consider the determination from \citet{2021MNRAS.508.3877G}. We also include in Figure \ref{ZZCetis} the ZZ Ceti stars extracted from the works of \citet{2016IBVS.6184....1B, 2017ApJS..232...23H, 2017ApJ...847...34S, 2019MNRAS.484.2711R, 2020AJ....160..252V, Romero22} (purple squares). All effective temperature and surface gravity values derived from spectroscopy were corrected by 3D convection \citep{tremblay, 2019A&ARv..27....7C}. 
As expected, most targets have masses close to the mean mass of white dwarfs $\sim 0.6~$M$_{\odot}$ \citep[e.g.][]{2021MNRAS.507.4646K, 2023MNRAS.518.3055O}. In our analyzed sample, there are five objects with stellar masses in the range of $0.30 \leq \mathrm{M}_* /\mathrm{M}_{\odot} \leq 0.45$, which correspond to low-mass white dwarfs \citep{Kilic2007, 2016A&A...595A..35I, Pelisoli2019} and can harbor either a He/C/O core or a He core, depending on the evolution of the progenitor star \citep{2022MNRAS.510..858R}. Also, one object, TIC~0264172524, has a photometric mass of $\sim 0.25$~M$_{\odot}$ likely an extremely low-mass (ELM) white dwarf variable.

\begin{table*}
\centering
 \caption{List of the 53 ZZ Ceti observed by TESS presented in this work. Column 1 indicates the TIC identifier. The sexagesimal right ascension (RA) and declination (DEC) in J2000 are listed in columns 3 and 4, and the G magnitude is listed in column 5. The effective temperature, $\log g$ and stellar mass determinations are listed in columns 6, 7 and 8, respectively. Column 9 indicates the reference for the atmospheric parameters. Data were taken from the works of (a) \citet{2020ApJ...898...84K} (b)
 \citet{2021MNRAS.508.3877G}, (c) \citet{2020MNRAS.497..130T} (d) \citet{2019MNRAS.482.4570G}, (e) Montreal White Dwarf Database \citep{MWDD} (f) \citet{2013AJ....145..136L} (g) \citet{2020AJ....160..252V} (h) S22 (i) \citet{2019MNRAS.486.2169K}, (j) \citet{2011ApJ...743..138G}, (k)  \citet{2017MNRAS.472.4173R} and (l) \citet{2012MNRAS.425.1394K}. The last column indicates whether the object was discovered as a ZZ Ceti by \citet{Romero22} (R22), or if it was confirmed as a pulsating DA white dwarf in this work (new). } 
   \begin{tabular}{lcccccccc}
    \hline
TIC & RA & DEC & G & $T_{\rm eff}$ [K] & $\log g$ & Mass [$M_{\odot}$]  &  Ref.  & \\
\hline
0001116746 & 102300.51 & +240704.5 & 16.44 & $11608\pm 82$ & $8.014\pm 0.013$ & $0.612\pm 0.007$ & a & new\\ 
$\cdots$   & $\cdots$  & $\cdots$  & $\cdots$ & $12127\pm 320$ & $8.043\pm 0.041$ & $0.629\pm 0.022$ & b & $\cdots$ \\ 
0007675859 & 181222.74 & +432107.6 & 16.24 & $12445\pm 142$ & $8.49\pm 0.009$ & $0.900 \pm 0.005$ & a & R22 \\
$\cdots$   & $\cdots$  & $\cdots$  & $\cdots$ & $12240\pm 214$ & $8.479\pm 0.023$ & $0.893 \pm 0.014$  & b & $\cdots$ \\ 
0014448610 & 072300.20 & +161704.4 & 15.08 & $11760\pm 80$ & $8.29\pm 0.02$ & $0.777\pm 0.012$ & c & new\\
0021187072 & 182606.04 & +482911.3 & 16.28 & $11808 \pm 228$ & $7.235\pm 0.025$ & $0.320\pm 0.006$ & b & R22 \\ 
0030545382 & 131630.82 & $-$392404.8 & 15.77 & $12093\pm 233$ & $7.221\pm 0.028$ & $0.3177\pm 0.007$ & b & new \\
0033717565 & 040536.39 & $-$762828.1 & 16.50 & $10675\pm 172$ & $7.639\pm 0.031$ & $0.447\pm 0.012$ & b & R22 \\ 
0055650407 & 045527.27 & $-$625844.6 & 14.97 & $11838\pm 150$ & $7.945\pm 0.019$ &  $0.576\pm 0.009$ & b & R22\\
0063281499 & 222858.15 & $-$310553.7 & 15.61 & $12200\pm 220$ & $8.02\pm 0.06$ & $0.616\pm 0.032$ &  k & R22 \\
$\cdots$   & $\cdots$  & $\cdots$  & $\cdots$ & $11712\pm 166$ & $7.981\pm 0.025$  & $0.594\pm 0.013$ & b & $\cdots$  \\
0072517198 & 134903.63 & +141954.9 & 16.64 & $10750\pm 350$ & $ 7.75\pm 0.038$ &  $0.493\pm 0.022$ & e & new \\
0079353860 & 211815.52 & $-$531322.7 & 15.92 & $11123\pm 44$ & $7.944\pm 0.004$ & $0.571\pm 0.002$  &  d & R22 \\ 
0081848974 & 111221.43 & $-$513003.9 & 16.40  & $11033\pm 191$ & $7.977\pm 0.035$ & $0.590\pm 0.018$ &  b & new \\
0088046487 & 095006.95 & +183733.9 & 16.68 & $11576\pm 298$ & $7.978\pm 0.050$ &  $0.592\pm 0.026$ & b & new\\ 
0094748632 & 073145.12 & +235353.0 & 16.14 & $11779\pm 222$ & $8.025\pm 0.032$ &  $0.618\pm 0.017$ & b & new\\ 
0103700861 & 102312.65 & +672506.5 & 16.80 & $8477\pm 103$ & $8.060\pm 0.033$ & $0.630\pm 0.018$ &  b & new\\ 
0114058447 & 034110.53 & +190800.4 & 15.84 & $9304\pm 106$ & $7.533\pm 0.030$ & $0.397\pm 0.011$ &  b & new\\ 
0141179495 & 120109.51 & +425947.4 & 16.11 & $11882\pm 179$ & $8.018\pm 0.026$ & $0.615\pm 0.014$ & b & new \\
0149863849 & 174349.28 & $-$390825.9 & 13.53 & $11604\pm 206$ & $8.087\pm 0.027$ & $0.652\pm 0.015$ & b & R22 \\
0159973152 & 032031.41 & $-$365657.6 & 16.50 & $10715\pm 193$ & $7.893\pm 0.042$ & $0.546\pm 0.021$ & b & new \\
0192937035 & 080856.78 & +461300.1 & 16.23 & $11542\pm 260$ & $7.398\pm 0.031$ & $0.363\pm 0.008$ &  b & new \\
0201860926 & 023156.01 & $-$535753.2 & 16.78 & $11353\pm 244$ & $7.895\pm 0.042$ & $0.548\pm 0.021$ & b & new \\
0230384389 & 190319.56 & +603552.6 & 15.01 & $11550\pm 178$ &  $8.07\pm 0.05$ & $0.642\pm 0.027$ & f & R22\\
$\cdots$   & $\cdots$  & $\cdots$ & $\cdots$ & $10858\pm 63$ &  $\cdots$        & $0.624\pm 0.006$ & g & $\cdots$\\ 
0261400271 & 061813.07 & $-$801155.2 & 14.90 & $12731\pm 762$ & $8.36\pm 0.04$ &  $0.819\pm 0.024$ & h & R22 \\
0264172524 & 193819.58 & +690911.3 & 16.42 & $13867\pm 467$ & $6.935\pm 0.047$ & $0.254\pm 0.006$ & b & new \\ 
0273206673 & 043350.99 & +485039.1 & 15.86 & $11433\pm 221$ & $7.966\pm 0.035$ & $0.586\pm  0.018$ & b & R22\\
0304024058 & 092256.24 & $-$681648.7 & 16.10  & $11368\pm 208$ & $7.955\pm 0.033$ &  $0.580\pm 0.017$ & b & R22\\ 
0317153172 & 232231.91 & $-$831313.3 & 16.47 & $11813\pm 314$ & $8.032\pm 0.042$ & $0.622\pm 0.022$ & b & R22 \\
0317620456 & 184335.47 & +274026.8 & 16.92  & $10566\pm 57$ &   $\cdots$  &  $0.603\pm 0.006$ & g & R22\\
0343296348 & 174344.00 & $-$742437.5 & 15.85 & $11597\pm 150$ & $7.968\pm 0.023$ &  $0.587\pm 0.011$ & b & R22 \\ 
0353727306 & 024029.66 & +663637.0 & 15.60 & $11554\pm 171$ & $8.020\pm 0.025$  & $0.615\pm 0.013$ & b & R22\\
0375199799 & 203750.60 & +061227.6 & 16.41 & $11310\pm 203$ & $8.007\pm 0.036$ &  $0.607\pm 0.019$ & b & new \\ 
0409732714 & 082403.35 & $-$121943.6 & 14.74 & $12176\pm 154$ & $7.874\pm 0.016$ &  $0.540\pm 0.007$ & b & new \\ 
0423658036 & 132924.81 & $-$235216.7 & 16.01 & $13910\pm 459$ & $8.02\pm 0.06$ & $0.620\pm 0.031$ & j & new \\
 $\cdots$   & $\cdots$  & $\cdots$ & $\cdots$ & $13650\pm 219$ & $8.002\pm 0.024$ & $0.610\pm 0.012$ & b & $\cdots$ \\
0453210132 & 075141.50 & +112029.8 & 16.41 & $11791\pm 62$ & $7.923\pm 0.009$ & $0.564\pm 0.005$ & a & new\\
 $\cdots$  & $\cdots$  & $\cdots$  & $\cdots$ & $11818\pm 175$ & $7.917\pm 0.028$ & $0.561\pm 0.014$ & b & $\cdots$ \\
0461203226 & 120650.89 & $-$380549.2 & 15.67 & $11866\pm 153$ & $8.073\pm 0.021$ & $0.645\pm 0.011$  & b & new \\
0600589802 & 010025.56 & +421840.9  & 16.56 & $11377\pm 256$ & $7.992\pm 0.044$ & $0.599\pm 0.023$ &  b & new \\
0631161222 & 012624.73 & $-$711711.9 & 16.96 & $11435\pm 362$ & $7.934\pm 0.059$ & $0.569\pm 0.030$ & b & R22 \\
0631344957 & 021328.27 & $-$643708.9 & 16.98 & $11574\pm 311$ & $7.995\pm 0.050$ & $0.601\pm 0.026$ & b & R22 \\
0640201450 & 032109.03 & +141056.0 & 17.17 & $11979\pm 488$ & $8.050\pm 0.070$ & $0.632\pm 0.037$ &  b & new\\
0661119673 & 044258.31 & +323715.6 & 17.37 & $10668\pm 345$ & $7.881\pm 0.082$ & $0.540\pm 0.032$ & b & R22\\
\hline
    \end{tabular}
    \label{tab:list1}
\end{table*}

\begin{table*}
\centering
 \caption{Continue from Table \ref{tab:list1}} 
   \begin{tabular}{lcccccccc}
    \hline
TIC & RA & DEC & G & $T_{\rm eff}$ [K] & $\log g$ & Mass [$M_{\odot}$]  &  Ref.  & \\
\hline
0762000503 & 072708.05 & +145216.2 & 17.22 & $11512\pm 339$ & $8.015\pm 0.061$ & $0.612\pm 0.032$ & b & new\\
0775564285 & 074613.81 & $-$302550.9 & 16.03 & $10891\pm  211$ & $7.938\pm 0.040$ &  $0.570\pm 0.020$ & b & new \\
0800126377 & 085128.17 & +060551.1 & 16.82 & $11427\pm 37$ & $8.128\pm 0.021$ & $0.675\pm 0.012$ & i & new\\
$\cdots$   & $\cdots$  & $\cdots$  & $\cdots$ & $11510\pm 227$ & $7.957\pm 0.040$ & $0.581\pm 0.020$ & b & $\cdots$ \\
0800420812 & 084055.71 & +130329.4 & 17.19 & $11562\pm 225$ & $7.981\pm 0.045$ & $0.594\pm 0.023$ & b & new\\ 
0804835539 & 085457.51 & $-$764621.9 & 16.91 & $15296\pm 476$ & $8.078\pm 0.046$ & $0.659\pm 0.051$ & b & R22 \\ 
0842451090 & 102132.75 & +140124.2 & 17.05 & $12191\pm 545$ & $7.841\pm 0.067$ & $0.525\pm 0.022$ & b & new\\
0900228144 & 110121.93 & +401546.1 & 17.47 & $11410\pm 560$ & $8.000\pm 0.110$ & $0.604\pm 0.058$ & b & new\\
0900762564 & 113738.38 & +801600.1 & 16.89 & $11939\pm 238$ & $7.831\pm 0.035$ & $0.524\pm 0.00$ & b & new\\ 
0902514572 & 114319.10 & -080115.4 & 17.13 & $11449\pm 259$ & $8.030\pm 0.050$ & $0.620\pm 0.027$ & b & new\\
1102242692 & 152809.27 & +553914.4 & 17.09 & $11180\pm 184$ & $7.86\pm 0.07$ & $0.531\pm 0.025$ & j & R22 \\ 
1102346472 & 145323.52 & +595056.2 & 17.16 & $12102\pm 345$ & $8.067\pm 0.045$ & $0.642\pm 0.024$ & b & R22\\
$\cdots$  & $\cdots$ & $\cdots$ & $\cdots$ & $11217\pm 250$ & $7.97\pm 0.038$ & $0.589\pm 0.05$ & i & $\cdots$ \\  
1860439362 & 192154.59 & +271437.4 & 17.00 & $11809\pm 294$ & $8.075\pm 0.044$ & $0.646\pm 0.024$ & b & new\\ 
1944049427 & 203804.29 & +234059.9 & 17.16 & $11338\pm 376$ & $8.020\pm 0.065$ & $0.614\pm 0.035$ & b & new \\ 
2045970633 & 234507.33 & +581315.0 & 16.88 & $11570\pm 302$ & $8.073\pm 0.049$ & $0.644\pm 0.027$ & b & new \\ 
\hline
    \end{tabular}
    \label{tab:list1-1}
\end{table*}

\section{TESS data}
\label{section3}

We downloaded the 2-min and 20-s cadence light curves for all known white dwarfs and candidates for white dwarfs \citep{2019MNRAS.482.4570G,2021MNRAS.508.3877G} brighter than G~$\leq$~17.5\,mag from The Mikulski Archive for Space Telescopes, hosted by the Space Telescope Science Institute (STScI)\footnote{http://archive.stsci.edu/} in FITS format. Data were processed on the basis of the Pre-Search data conditioning pipeline \citep{2016SPIE.9913E..3EJ}. The times and fluxes (PDCSAP FLUX) were extracted from the FITS
files, with the times given in barycentric corrected dynamical Julian days \citep[BJD – 2457000, corrected for leap seconds, see][]{2010PASP..122..935E}. The fluxes were converted to differential flux $\Delta I/I$, and then transformed to amplitudes in parts per thousand (ppt), which corresponds to the millimodulation amplitude unit (mma)\footnote{1 mma= 1/1.086 mmag= 0.1\% = 1 ppt; see, e.g., \citet{2016IBVS.6184....1B}.}. Data were sigma-clipped at 5$\sigma$ to remove outliers that depart from the median by 5$\sigma$. 

We computed their Fourier transforms (FTs) looking for signatures of pulsations or binarity
above the 1/1000 false-alarm probability (FAP). The FAP was calculated by reshuffling the data 1000 times, maintaining the same time base, and computing the Fourier transform, selecting the highest peak. For prewhitening, we used a nonlinear least-squares (NLLS) method, by simultaneously fitting each pulsation frequency in a waveform
$A_i\sin(\omega_1 t + \phi)$, with $\omega = 2\pi / P$ and $P$ the period. This iterative process was performed starting with the highest amplitude peak until no peak appeared above the false alarm probability significance threshold of 0.1\%. For objects showing only one variation period, we ask for an amplitude above 10 \% above the FAP detection limit. We analyze the concatenated light curve from different sectors when observed, using the \textsc{TESS-LS} script.

The flux corresponding to the white dwarf ranged from CROWDSAP = 0.021--0.985 due to the large pixel scale of the TESS plates, which means that the total flux from the white dwarf in the extracted aperture ranged from 2.1\% to 98.5\%. To confirm that the variations are from the white dwarf, we checked all stars around 120"x120" in Gaia EDR3 for other possible variables or parallax and proper motion companions. The values for the PDCSAP flux were corrected for crowding using the CROWDSAP value; thus, the reported amplitudes were corrected for flux dilution.

\section{Periods and data analysis}
\label{section4}

In this section, we present the results from the light curve analysis of all objects presented in Table~\ref{tab:list1}, including 32 new bright ZZ Ceti stars. The period list for each object is presented in Tables~\ref{tab:list2} and \ref{old} for the new ZZ Cetis, and the pulsating DA white dwarf stars reported by \citet{Romero22}, respectively. We also include the sectors where each object was observed and the amplitude detection limit for the false alarm probability FAP=1/1000. 

\begin{figure*}
	\includegraphics[width=0.95\textwidth]{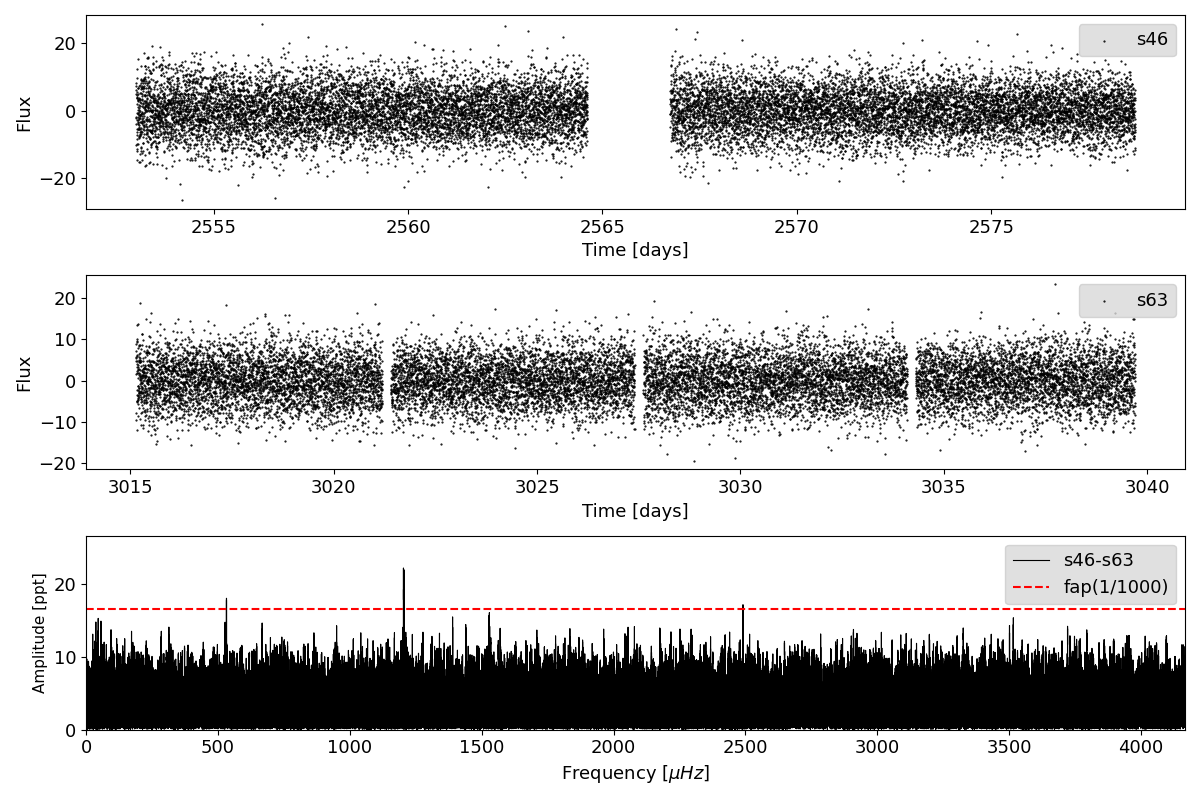}
    \caption{Light curves from sectors 46 (120s--cadence) and 63 (20s--cadence) for TIC~0902514572, shown in the top and middle panel, respectively. The bottom panel presents the Fourier transform of the concatenated data. The horizontal red line corresponds to the false-alarm probability FAP=1/1000 detection limit.}
    \label{00902514572}
\end{figure*}

From the data obtained by TESS in Cycles 4 and 5, we found 20 new ZZ Ceti stars. As an example, we show the light curves and FT for TIC~0902514572 in Figure~\ref{00902514572}. The top panel shows the light curve for the data from Sector~46 with 120\,s cadence, while the middle panel shows the data from Sector~63 with 20\,s cadence. The bottom panel shows the FT for the concatenated data from both sectors, where there are three distinctive peaks above the detection limit of FAP (1/1000), indicated as a horizontal line at 16.60~ppt. The main peak is located at a frequency of $\sim 1200\,\mu$Hz, corresponding to a period $\sim 830$~s. Two objects were previously indicated as DAV stars. TIC~0014448610 was reported as a DAV pulsator by \citet{2020AJ....160..252V} with a detected period at 490~s. Using data from Sectors 44 to 46 we detected a total of 6 periods. Finally, \citet{2019MNRAS.486.4574R} reported the detection of one period at $\sim 490$~s for TIC~0800420812, while we detect two periods at 507.76 and 333.75~s.

For 12 objects, the data from the first three cycles, Sectors 1 to 39, lead to no pulsation-related variability detection, with peaks below the FAP(1/1000). However, new data from Cycles 4 and 5, particularly with 20\,s cadence, led to the detection of pulsation variability in these objects.

\begin{table*}
    \centering
        \caption{Detected periods for the 32 new ZZ Cetis from TESS. For each object, we list the sectors where the target was observed by TESS, indicating the 20~s cadence runs with "f'' (Column~2), the value of the amplitude detection limit for false-alarm probability FAP(1/1000) (Column~3), and the list of periods compatible with stellar pulsations in white dwarfs (Column~4). We truncate all periods to two decimal places only because the uncertainties in the theoretical models are of the order of 1~s. }
    \begin{tabular}{cccc}
    \hline
    TIC	  &   Sector(s)		& FAP(1/1000) [ppt] & $\Pi$ [s] (A [ppt])  \\
    \hline
0001116746 & f48 & 7.24 & 303.01 (8.61) \\ 
0014448610 & f44-46 & 2.84 & 472.13 (5.42), 496.91 (4.89), 634.34 (4.70), \\
$\cdots$   & $\cdots$ & $\cdots$ &  679.27 (3.47), 527.92 (3.08), 373.57 (3.04) \\ 
0030545382 & 11,f37,f64 & 3.59 & 1160.54 (9.82), 781.18 (8.89), 718.18 (5.15), 1141.29 (4.90) \\ 
0072517198 & f50 & 8.73 & 848.91 (14.28) \\ 
0081848974 & f36-f37,f63-f64 & 7.01 & 874.58 (9.52) \\
0088046487 & f45 & 5.62 & 632.15 (11.28), 885.76 (8.64), 829.85 (6.20) \\
0094748632 & f44-f46 & 3.86 & 306.97 (6.28), 257.63 (5.69) \\
0103700861 & f40-f41,f47 & 4.50 & 536.92 (5.54), 715.90 (4.59) \\
0114058447 & f42-f44 & 2.73 & 1402.65 (6.70), 1401.00 (3.28), 319.91 (2.92) \\
0141179495 & 22,f49 & 3.87 & 253.86 (4.46) \\
0159973152 & 31 & 5.444 & 977.55 (7.27),  922.33 (6.43), 1023.48 (6.07) \\
0192937035 & 20,47,f60 & 4.55 & 298.27 (5.25) \\
0201860926 & 29-30,69 & 4.40 & 893.74 (6.82), 880.74 (7.31), 525.06 (5.89), \\ 
   $\cdots$  & $\cdots$ & $\cdots$  & 522.77 (4.89), 520.42 (7.53), 486.05 (4.49) \\
0264172524 & 14-16,18-26,f40-f41,f47-f51,f53-f60 & 2.45 & 463.08 (5.69), 336.26 (2.78), 847.81 (2.62) \\ 
0375199799 & f55 & 9.17 & 328.12 (9.56),305.53 (12.98) \\ 
0409732714 & 34,f61 & 1.34 & 448.23 (1.51), 940.48 (1.26) \\
0423658036 & 10,f37,f64 & 3.637 & 372.32 (4.344) \\
0453210132 & 34,f44-f46 & 5.85 & 200.31 (6.93) \\
0461203226 & f37,f64 & 3.73 & 297.78 (4.39), 259.91 (10.29) \\ 
0600589802 & f57 & 8.14 & 959.02 (25.88), 966.99 (8.87), 868.64 (17.20), \\
 $\cdots$  & $\cdots$ & $\cdots$  &  530.15 (12.13), 479.61 (8.37) \\
0640201450 & 42-44 & 11.03 & 254.87 (13.59) \\ 
0762000503 & 44-46 & 10.04 & 453.66 (15.26), 299.20 (13.70), 1427.68 (10.18)\\
0775564285 & f61 & 8.25 & 852.96 (12.34), 895.81 (9.04), 627.84 (10.19) \\
0800126377 & f44-f45,61 & 8.14 & 325.05 (10.18), 326.31 (9.26) \\
0800420812 & 44-46 & 10.87 & 507.76 (23.31), 333.75 (10.84) \\
0842451090 & 45-46 & 7.42 & 311.80 (10.51) \\
0900228144 & 48 & 15.99 & 578.17 (29.82), 826.54 (28.02), 542.77 (18.81) \\
0900762564 & f40-41,47,f53,f60 & 4.32 & 268.39 (5.12), 260.78 (4.61), 885.55 (4.33) \\
0902514572 & 46,f63 & 16.60 & 831.13 (22.21), 829.56 (21.94), 1879.21 (18.11), 401.51 (17.24)\\
1860439362 & 41,54 & 43.86 & 575.26 (88.17), 461.79 (67.71) \\
1944049427 & 41,55 & 43.83 & 606.27 (85.79)\\
2045970633 & 17,24,f57-f58 & 26.43 & 883.61 (28.82), 930.02 (27.01), 726.49 (26.10) \\
\hline
    \end{tabular}
    \label{tab:list2}
\end{table*}

From the sample of 74 new ZZ Cetis observed in the first three years of the TESS mission, analyzed by \citet{Romero22}, there are 28 objects that were reobserved during the fourth and fifth cycles. Of these 28 targets, 21 of them showed a change in their period list. These 21 objects are listed in Table~\ref{old}, where we also list the periods detected from the FT considering all observed sectors. 

\begin{figure*}
	\includegraphics[width=0.95\textwidth]{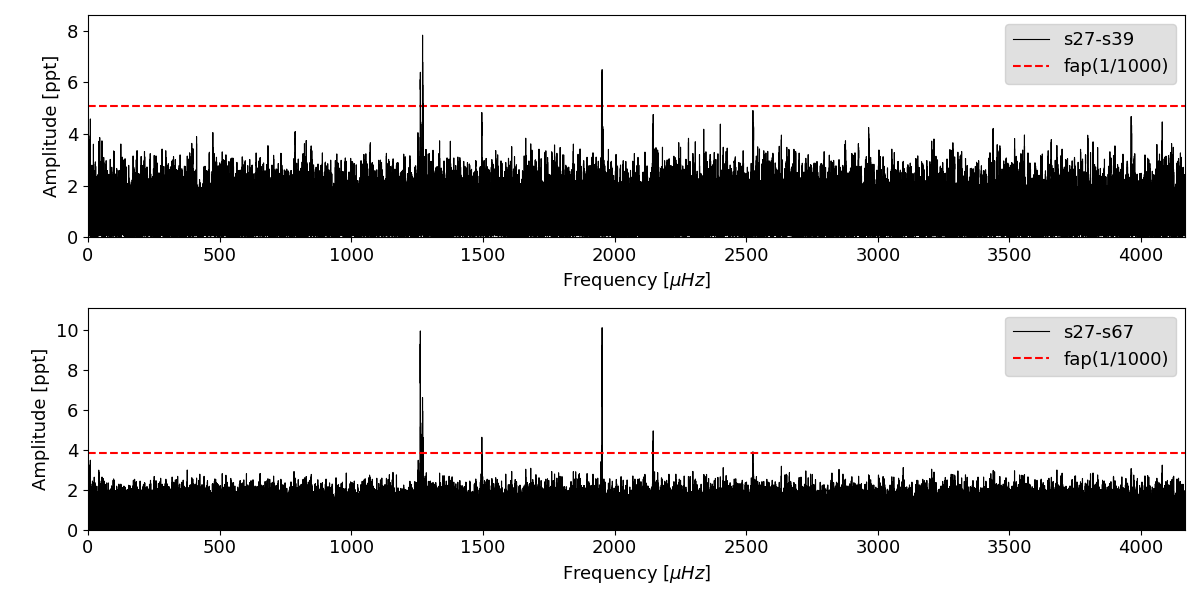}
    \caption{Fourier transform for the data for TIC~0317153172 data for sectors 27 to 39 (top panel), and for sectors 27 to 67 (bottom panel). The horizontal red line corresponds to the false-alarm probability FAP=1/1000 detection limit.  }
    \label{317153172}
\end{figure*}

For most objects, the additional data lead to the detection of more periods, highlighted in italic in Table~\ref{old}. For instance, for TIC~0317153172, three new low-amplitude periods were detected after data from Sectors 66 and 67 were included in the Fourier transform analysis. Figure~\ref{317153172} shows the FT for  TIC~0317153172 from Sectors 27 and 39 on the top panel and for all sectors on the bottom panel. Note that three new peaks in the FT are above the FAP(1/1000) detection limit when data from the two sectors from Cycle 5 are included. In other cases, the additional periods correspond to components of rotational multiplets that were not resolved with the previous data. For instance, for TIC~1102242692, three components of a multiplet centered in 1009.03~s are above the detection limit, when 20-s-cadence data from Sectors $49-51$ are considered. 

For three objects, there is no observed variability above the detection limit after later reobservation by TESS. TIC~0261400271 was reported to be variable by \citet{Romero22}, showing two short periods of $\sim 300$s and $\sim 380$ s, with low amplitudes close to the detection limit. These periods are not present in the Fourier transform when 20-s data from Sectors 61, 62, and 64 are added to those from previous sectors; thus, we label this object as NOV in Table~\ref{old}. Furthermore, \citet{Romero22} reported the detection of a peak from the FT of TIC~0804835539, and photometric variability was reported for TIC~0317620456 by \citet{2020AJ....160..252V} and \citet{Romero22}. However, we found that the variability of these objects was coming from a nearby object in the large TESS pixels and not from the white dwarf; thus, they are also labeled NOV.  

\begin{table*}
\centering
	\caption{List of stars identified as ZZ Cetis by \citet{Romero22} using data from Sector 1 to Sector 39, showing a change in the period list when data from Sectors 40 to 69 are added to the analysis.  We list TIC (column 1), all sectors where they were observed (column 2), the detection limit FAP(1/1000) (column 3), and all detected periods (column 4). The periods detected after the new data were added to the FT are in italics.  We truncate all periods to two decimals places only because the uncertainties in the theoretical models are of the order of 1~s. }
	\label{old}
	\begin{tabular}{rccc} 
   \hline
    TIC	  &   Sector(s)		& FAP(1/1000) [ppt] & $\Pi$ [s] (A [ppt])  \\
    \hline
0007675859 & 25,26,f40,f52-f54 & 4.09 & 798.65 (4.72), 743.46 (4.30), {\it 575.41 (5.87)}, 356.10 (5.26), \\ 
$\cdots$   &  $\cdots$      & $\cdots$  & 353.25 (10.5), {\it 254.49 (4.14)} \\
0021187072 & 25,26,f40-f41,f52-f54 & 3.13 & 1076.93 (4.13), {\it 887.47 (3.32)}, {\it 823.67 (3.25)}, {\it 794.27 (3.13)}, \\
  $\cdots$  &  $\cdots$ &  $\cdots$       & {\it 776.71 (3.56)} \\ 
0033717565 & 27-29,32,35-36,39,f61-f63,f65-f68 & 1.71 & 364.92 (9.30), 526.97 (6.08), {\it 243.02 (2.04)}, {\it 703.78 (1.74)}, \\ 
 $\cdots$  &     $\cdots$    &   $\cdots$  & {\it 615.13 (1.74)}, {\it 484.17 (1.76)}, {\it 460.54 (1.77)}\\
0055650407 & 11-13,f27-f39,f61-f64 & 0.36 & 320.76 (1.59), {\it 318.01 (0.59)}, 262.46 (7.19), {\it 261.61 (0.47)}, \\
 $\cdots$  & $\cdots$ & $\cdots$ & {\it 206.70 (0.39)}, {\it 200.66 (0.88)}, 200.08 (4.42), {\it 199.52 (0.59},\\
 $\cdots$  &     $\cdots$    &   $\cdots$  &  {\it 153.26 (0.41)}, {\it 127.03 (0.48)}, 126.84 (1.80) \\
0063281499 & 01,f28,f68 & 2.071 & 383.70 (2.73), 320.51 (7.40), {\it 269.13 (2.31)} \\ 
0079353860 & 01, 27, 67 & 2.13 & {\it 982.02 (2.34)}, 945.15 (2.63), 525,55 (2.37) \\  
0149863849 & f39,f66 & 1.49 & {\it 835.84 (2.08)}, {\it 811.00 (1.58)}, {\it 804.01 (1.51)}, {\it 800.00 (3.14)}, \\ 
$\cdots$ & $\cdots$ & $\cdots$ & {\it 796.04 (1.53)}, {\it 789.27 (3.27)} , {\it 769.67 (1.99)}, {\it 763.40 (3.48)}, \\ 
$\cdots$ & $\cdots$ & $\cdots$ & {\it 759.78 (9.91)}, {\it 756.19 (2.80)}, {\it 750.10 (4.59)}, {\it 660.62 (2.43) }, \\
$\cdots$ & $\cdots$ & $\cdots$ & 568.09 (1.96), 491.21 (4.09), 487.35 (1.94), 398.96 (10.3) {\it 389.68 (2.39)}, \\
$\cdots$ & $\cdots$ & $\cdots$ & {\it 379.89 (3.30)}, {\it 261.18 (1.69)}\\
0230384389 & 14-26,f40-f41,f47-f54,f56-f60 & 0.47 & {\it 756.81 (0.71) }, {\it 753.07 (0.85)}, 749.62 (0.74),  707.93 (1.41), \\ 
 $\cdots$  &   $\cdots$  & $\cdots$  & {\it 494.78 (0.90)}, 493.86 (0.85), 457.17 (1.32), 1633.58 (0.59) \\
0261400271 & 1,4,7,8,11-13,f27-f28,f31,f34, & 0.49 & NOV \\ 
$\cdots$  & f38-f39,f61-f62,f64 & $\cdots$ & $\cdots$ \\ 
0273206673 & 19,f59 & 8.02 & {\it 941.83 (8.20)}, {\it 914.81 (12.60)}, 892.97 (10.01), 874.55 (7.44),  \\ 
$\cdots$   &  $\cdots$ & $\cdots$ & 844.08 (9.07), 827.16 (16.81), {\it 826.47 (14.44)}, {\it 825.31 (9.96)}, \\
 $\cdots$  & $\cdots$  &  $\cdots$ & 746.68 (10.26), 698.90 (10.46), 688.96 (8.06),  663.25 (8.41),  \\
 $\cdots$  & $\cdots$ &  $\cdots$  & 583.44 (17.43), 511.41 (10.91), 464.42 (9.59) \\
0304024058 & 10-11,f36-f38,f63-f64 & 3.54 & {\it 740.98 (3.73)}, 623.29 (5.49), 579.47 (13.05), 506.20 (11.54), \\ 
  $\cdots$  & $\cdots$ &  $\cdots$  &  400.28 (12.32)\\ 
0317153172 & 27,f39,f66-67 & 3.81 & 792.07 (9.94), 786.89 (6.59), {\it 668.39 (4.63)}, 512.05 (10.01),  \\ 
  $\cdots$  & $\cdots$ &  $\cdots$  & {\it 465.78 (4.98)}, {\it 395.94 (3.89)}\\
0317620456 & 26,f40,f53-f54 & 4.16 & NOV \\ 
0343296348 & 12-13,f39,f66 & 4.41 & 288.27 (10.81), {\it 287.26 (7.11)} \\ 
0353727306 & 18-19,25,f52,f58-f59 & 5.00 & {\it 876.08 (8.10)}, {\it 873.87 (8.94)}, {\it 771.30 (5.22)}, {\it 750.84 (5.49)}, \\ 
$\cdots$   & $\cdots$           & $\cdots$ & {\it 729.84 (5.48)}, 545.82 (22.0), 470.24 (9.35), 463.54 (5.70), \\
 $\cdots$  &  $\cdots$          & $\cdots$ &  404.84 (8.54), {\it 399.56 (5.07)} \\
0631161222& 27,29,f68 & 5.59 & 707.97 (15.80), 679.82 (26.27), 466.62 (10.29), 403.63 (12.45), \\ 
 $\cdots$  &  $\cdots$  & $\cdots$  & 367.44 (8.35), {\it 346.79 (6.16)} \\
0631344957 & 28-29,f68 & 6.23 & 363.14 (7.94), {\it 313.44 (6.44)} \\ 
0661119673 & 19,43-44,f59 & 18.34 & 626.37 (46.78), {\it 2$f_1=$313.10 (19.11)} \\ 
0804835539 & 37-39,f64-f66 & 6.04 & NOV \\ 
1102242692 & 16,22,24,f49-f51 & 4.89 & {\it 1015.04 (5.73)}, 1009.027 (8.37), {\it 1003.09 (4.94) }\\ 
1102346472 & 15-16,22-23,48-50 & 6.93 & 458.13 (23.45), {\it 1489.65 (7.27)} \\ 
\hline\hline
\end{tabular}
\end{table*}

\section{Asteroseismological analysis}
\label{section5}

For the asteroseismological fit, we analyzed the detected period list shown in Tables~\ref{tab:list2} and \ref{old}. Periods identified as components due to rotation, harmonics, and linear combinations are not taken into account in the asteroseismological fit, since we only consider $m=0$ components in the computations of the theoretical periods. Rotational splitting is used to estimate rotation periods after a best-fit model is found.  

For our seismological fit, we employ an updated version of the DA white dwarf model grid used in \citet{Romero22}.  DA white dwarf models were obtained from fully evolutionary computations using the stellar evolution code LPCODE \citep[see][for details]{2010ApJ...717..897A,2010ApJ...717..183R,2015MNRAS.450.3708R}. The computations start in the zero-age main sequence, going through the central hydrogen and helium burning stages and the giant phases, where the mass loss episodes occur. In particular, we compute the Thermally Pulsing AGB phase, where most of the mass is lost to the interstellar medium. Finally, after the evolution during the post-AGB phase, where the effective temperature increases rapidly at almost constant luminosity, the computations arrive at the white dwarf cooling sequence. We computed the evolution through the cooling curve, including the DA instability strip between $\sim 12\, 500$ and $\sim 10\, 000$ K.

The first version of the model grid was presented in \citet{2012MNRAS.420.1462R}, consisting of C/O-core white dwarf models with 11 different stellar masses between 0.525~M$_{\odot}$ and 0.878~M$_{\odot}$, and 6 to 8 different values of the hydrogen envelope mass, depending on the stellar mass. The grid was extended to higher masses, up to 1.05~M$_{\odot}$ by \citet{2013ApJ...779...58R}, who added 6 additional sequences. In further work, the grid was extended by adding models with different stellar masses and hydrogen envelopes \citep{2017ApJ...851...60R, 2019MNRAS.490.1803R}. The current model grid is composed of 30 stellar mass values between 0.493~M$_{\odot}$ to 1.05~M$_{\odot}$, with a hydrogen envelope mass in the range of $\sim 3.5\times 10^{-4}$ to $\sim 10^{-10}$~M$_*$, adding to a total of 611 cooling sequences.

\begin{table}
	\centering
	\caption{Main Characteristics of DA WD Models Set}
	\label{tab:grid}
	\begin{tabular}{cccc} 
\hline
$M_*/M_{\odot}$ & $-\log (M_H/M_*)$ & $-\log (M_{He}/M_*)$ & $X_O$ \\
\hline
0.493 & 3.50 & 1.08 & 0.720 \\
0.510 & 3.53 & 1.19 & 0.703 \\
0.525 & 3.62 & 1.31 & 0.709 \\
0.534 & 3.65 & 1.38 & 0.677 \\
0.542 & 3.68 & 1.35 & 0.685 \\
0.548 & 3.74 & 1.38 & 0.697\\
0.550 & 3.65 & 1.39 & 0.698 \\
0.560 & 3.70 & 1.42 & 0.691\\
0.570 & 3.82 & 1.46 & 0.696\\
0.580 & 3.86 & 1.57 & 0.698\\
0.593 & 3.93 & 1.62 & 0.704\\
0.609 & 4.02 & 1.61 & 0.723\\
0.621 & 4.04 & 1.68 & 0.732 \\
0.632 & 4.25 & 1.76 & 0.755 \\
0.646 & 4.12 & 1.83 & 0.742\\
0.660 & 4.26 & 1.92 & 0.730\\
0.674 & 4.35 & 1.97 & 0.707\\
0.686 & 4.35 & 2.03 & 0.711\\
0.690 & 4.46 & 2.04 & 0.684\\
0.705 & 4.45 & 2.12 & 0.661 \\
0.721 & 4.50 & 2.14 & 0.659\\
0.745 & 4.62 & 2.18 & 0.657\\
0.771 & 4.70 & 2.23 & 0.655 \\
0.800 & 4.84 & 2.33 & 0.648\\
0.820 & 4.93 & 2.41 & 0.639 \\
0.837 & 5.00 & 2.50 & 0.640\\
0.860 & 5.08 & 2.55 & 0.624 \\
0.878 & 5.07 & 2.59 & 0.611\\
0.900 & 5.28 & 2.72 & 0.609\\
0.917 & 5.41 & 2.88 & 0.609\\
0.949 & 5.51 & 2.92 & 0.614\\
0.976 & 5.68 & 2.96 & 0.613\\
0.998 & 5.70 & 3.11 & 0.629\\
1.024 & 5.74 & 3.25 & 0.631\\
1.050 & 5.84 & 2.96 & 0.613\\
\hline
\end{tabular} 
\end{table}

For hydrogen envelopes thinner than $10^{-10}$,$M_H/M_*$ it is expected that the outer convective zone will mix hydrogen into the more massive helium layer before reaching the ZZ Ceti instability strip, turning the star into a DB white dwarf \citep{2020MNRAS.492.5003O, 2020MNRAS.492.3540C}. The stellar mass values of our complete model grid are listed in Column~1 of Table~\ref{tab:grid}, together with the hydrogen (Column~2) and helium (Column~3) content as predicted by standard stellar evolution, and the central oxygen abundance by mass ($X_O$) in Column~4.
 
Nonradial adiabatic g-mode pulsations were computed for models with effective temperatures from $13\,500$ K to $9\,500$~K, using the adiabatic version of the \textsc{LP-PUL} pulsation code \citep[see][for details]{2006A&A...454..863C}. We consider dipolar ($\ell=1$) and quadrupolar ($\ell=2$) modes with periods up to 2000~s.  

For each object, we searched for an asteroseismologically representative model that best matches the observed periods. To this end, we seek for the theoretical model that minimizes the quality function,
\begin{equation}
    \chi^2(M_*, M_{\rm H}, T_{\rm eff})=\frac{1}{N-1}\sqrt{\sum_{i=1}^N{\rm min}\left[\Pi_i^{\rm th}-\Pi_i^{\rm obs} \right]^2},
\end{equation}
where $N$ is the number of observed periods, $\Pi^{\rm th}_i$ is the theoretical period that best fits the observed period $\Pi^{\rm obs}_i$ in the model. When the star only shows one period, we set the factor $1/(N-1)$ to one. For the period fit, we consider only periods identified as $m=0$ modes. Thus, we do not consider periods corresponding to harmonics or linear combinations, or other components from rotational multiplets.  
For stars with a spectroscopic or photometric mass below the minimum value of our C/O-core grid (0.493~M$_{\odot}$) we also perform a preliminary asteroseismological fit with He--core white dwarf models with stellar masses from 0.17 to 0.45~M$_{\odot}$ \citep{2012A&A...547A..96C}, considering only thick hydrogen envelopes, as predicted by the previous evolution, and $\ell =1$ modes. 
External constraints are considered in the fitting procedure, especially for objects that show only one or two periods. For example, the values of the effective temperature and stellar mass determinations in Table~\ref{tab:list1} are taken into account as additional restrictions. However, note that some particular short periods ($\lesssim$ 200\,s) can be strong constraints on their own, since they propagate in the inner parts of the star and are particularly sensitive to the inner structure.

The results of our asteroseismological fits are presented in Table~\ref{tab:list1-res} for all objects analyzed in this work. The values for the stellar mass, the thickness of the hydrogen envelope, and the effective temperature for the seismological model are listed in columns 2, 3, and 4, respectively. In column 5 we list the values for the theoretical periods that better fit the observed periods, along with the corresponding harmonic degree $\ell$ and the radial order $k$. The value of the quality function $\chi^2$ is shown in column 9 for objects showing more than one period. The first model listed is the one we choose to be the best-fitting model for that particular object.

\begin{table*}
\centering
\caption{Best fit model for the 50 ZZ Cetis analyzed in this work using the list of observed modes. The stellar mass, hydrogen envelope, and effective temperature are listed in columns 2, 3, and 4, respectively. We list the theoretical periods in Column~5, along with the harmonic degree and the radial order. The value of the quality function $\chi^2$ in seconds is listed in Column~6, for objects with more than one period. }
\begin{tabular}{lccccc}
\hline
TIC & M/M$_{\odot}$ & $-\log($M$_{\rm H}$/M$_*)$ & T$_{\rm eff}$ [K] & $\Pi$[s] ($\ell, k$) & $\chi^2$ [s] \\
\hline
0001116746 & 0.609 & 4.02 & 12080 & 303.02 (1,5) & $\cdots$\\ 
0007675859 & 0.900 & 8.37 & 12180 & 355.26 (1,6), 575.22 (1,12), 799.07 (1,18), 743.19 (2,30), 253.69 (1,45) & 0.58 \\
0014448610 & 0.609 & 4.97 & 11510 & 428.67 (1,7), 492.59 (1,9), 634.65 (2,23) & 0.96\\
$\cdots$ & $\cdots$ & $\cdots$ &  $\cdots$ & 676.63 (2,25), 531.04 (2,19), 371.74 (2,12) &  $\cdots$\\
0021187072 & 0.542 & 5.43 & 11740 & 1077.96 (1,19), 884.93 (2,27), 824.14 (2,24), 776.69 (1,13) & 0.73 \\
0030545382 & 0.534 & 6.83 & 11800 & 1161.46 (1,19), 780.01 (1,12), 718.35 (1,11), 1137.87 (2,33) & 1.25 \\
0033717565 & 0.570 & 8.31 & 11600 & 366.26 (1,4), 524.82 (1,7), 241.38 (2,5), 458.28 (2,12) & 0.75 \\
 $\cdots$ & $\cdots$ & $\cdots$ & $\cdots$ & 482.95 (2,13), 616.99 (2,17), 704.82 (2,20) & \\  
0055650407 & 0.570 & 3.92 & 12380 & 263.94 (1,3), 204.97 (2,5), 125.53 (1,1), 319.56 (1,5), 153.35 (2,3) & 1.35 \\
 $\cdots$  & 0.593 & 3.93 & 11630 & 262.18 (1,3), 193.95 (1,2), 121.80 (1,1), 314.04 (1,5),152.15 (2,3) & 2.62\\
0063281499 & 0.570 & 7.83 & 12000 & 319.05 (1,3), 384.69 (1,50, 269.22 91,2)  &  0.60 \\    
0072517198 & 0.510 & 6.33 & 11000 & 848.92 (1,3) & $\cdots$\\
0079353860 & 0.690 & 8.38 & 11370 & 981.62 (1,18), 945.68 (1,17), 526.50 (1.8) & 0.56\\
0081848974 & 0.570 & 5.47 & 11120 & 874.56 (1,15) & $\cdots$ \\
0088046487 & 0.660 & 4.25 & 11370 & 632.33 (1,13), 885.36 (1,19), 830.12 (2.32) & 0.25 \\
0094748632 & 0.593 & 3.93 & 12350 & 306.99 (1,5), 257.68 (1,3) & 0.01 \\ 
0103700861 & 0.705 & 7.35 & 10190 & 537.03 (1,8), 715.98 (1,12) & 0.11 \\
0114058447 & 0.593 & 7.34 & 11520 & 1400.80 (1,25), 320.31 (1,4)  & 0.45 \\
 $\cdots$  & 0.358 & 3.18 &  9470 & 1400.03 (1,21), 335.90 (1,3) & 16.02 \\  
0141179495 & 0.632 & 6.34 & 11690 & 253.86 (1,3) & $\cdots$ \\
0149863849 & 0.660 & 4.24 & 11590 & 839.63 (1,18), 798.34 (1,17), 760.54 (1,16), 658.46(2,25), & 1.06 \\
$\cdots$  & $\cdots$ & $\cdots$ & $\cdots$ & 568.16 (1,11), 492.30 (2,18), 486.85 (1,10), 402.11 (1,7), 383.53 (2,13) & $\cdots$ \\
0159973152 & 0.542 & 6.83 & 10450 & 977.63 (1,15), 922.01 (2,25), 1023.50 (2,28), 0.17 \\
$\cdots$   & 0.570 & 9.33 & 11210 & 976.71 (1,15), 921.76 (1,14), 1024.25 (1,16) & 0.64 \\
0192937035 & 0.550 & 3.65 & 11830 & 298.22 (1,4) &  $\cdots$\\
$\cdots$   & 0.450 & 3.43 & 11640 & 298.18 (1,3)  &  $\cdots$\\
0201860926 & 0.534 & 6.13 & 11280 & 879.13 (1,14), 522.82 (1,7), 486.32 (2,13) & 0.81 \\
0230384389 & 0.660 & 5.85 & 11660 & 707.50 (1,13), 456.69 (1,7), 493.99 (1,8), 753.78 (1,14), 1635.17 (2,56) & 0.51\\
0264172524 & 0.686 & 5.55 & 12300 & 462.85 (1,8), 337.44 (1,5), 848.23 (1,17) & 0.64\\
0273206673 & 0.705 & 5.26 &11280 & 584.85 (1,11), 699.51 (1,14), 746.95 (1,15), 829.54 (1,17), & 0.75\\
$\cdots$  & $\cdots$ & $\cdots$ & $\cdots$ & 463.45 (2,16), 509.82 (2,18), 664.26 (2,24), 683.93 (2,25), &  $\cdots$\\
$\cdots$  & $\cdots$ &  $\cdots$ & $\cdots$ & 845.79 (2,31), 888.78 (2,33), 916.13 (2,34), 944,73 (2,35)  &  $\cdots$\\
$\cdots$ & 0.675 & 4.87 & 11310 & 587.57 (1,11), 698.45 (1,14), 748.89 (1,15), 831.58 (1,17), & 0.86 \\
$\cdots$ & $\cdots$ & $\cdots$ &  $\cdots$ & 464.03 (2,16), 510.09 (2,18), 664.03 (2,24), 690.29 (2,25), &  $\cdots$\\
$\cdots$ & $\cdots$ & $\cdots$ & $\cdots$ & 837.84 (2,31), 891.14 (2,33), 916.11 (2,34), 941.67 (2,35) &  $\cdots$\\  
0304024058 & 0.609 & 4.95 & 11210 & 581.61 (1,10), 396.42 (1,6), 505.12 (1,8), 624.53 (1,11), 738.84 (2,25 & 1.29\\
$\cdots$ & 0.593 & 6.33 & 11360 & 579.01 (1,9), 401.61 (2,11), 507.23 (2,15), 624.14 (2,19), 739.53 (2,23) & 0.61 \\
0317153172 & 0.686 & 4.43 & 11920 & 794.04 (1,17), 787.36 (2,30), 668.27 (1,14), 511.24 (1,10), & 0.44\\
$\cdots$ & $\cdots$ & $\cdots$ & $\cdots$ & 464.86 (1,14), 395.88 (1,7) & $\cdots$\\
0343296348 & 0.548 & 4.27 & 11310 & 287.77 (1,3) & $\cdots$\\
0353727306 & 0.621 & 4.04 & 11470 & 872.48 (1,18), 775.95 (2,29), 752.41 (2,28), 729.70 (1,15) & 1.71 \\
$\cdots$ & $\cdots$ & $\cdots$ & $\cdots$ & 538.20 (1,10), 467.11 (1,8), 408.31 (1,7) & $\cdots$ \\
$\cdots$ & 0.690 & 6.25 & 11310 & 874.43 92,30), 765.15 (2,26), 751.76 (1,14), 733.20 (2,25), & 1.24 \\
$\cdots$ & $\cdots$ & $\cdots$ & $\cdots$ & 543.99 (1,9), 468.88 (2,15), 404,66 (2,15) & $\cdots$ \\
0375199799 & 0.570 & 4.08 & 11810 & 305.21 (1,4), 328.12 (1,5) & 0.34 \\
0409732714 & 0.542 & 6.23 & 12200 & 448.23 (1,6) &  $\cdots$\\
0423658036 & 0.623 & 8.33 & 13700 & 372.31 (1,5) &   $\cdots$\\
$\cdots$ & $\cdots$ & $\cdots$ & $\cdots$ &  568.16 (1,11), 486.85 (1,10), 492.30 (2,18), 402.12 (1,7), 383.53 (2,13) & \\   
\hline
\end{tabular}
\label{tab:list1-res}
\end{table*}

\begin{table*}
\centering
\caption{Continue from Table \ref{tab:list1-res} }
\begin{tabular}{lccccc}
\hline
TIC & M/M$_{\odot}$ & $-\log($M$_{\rm H}$/M$_*)$ & T$_{\rm eff}$ [K] & $\Pi$[s] ($\ell, k$) & $\chi^2$ [s] \\
\hline
0453210132 & 0.593 & 5.64 & 12100 & 200.34 (1,2) & $\cdots$\\
0461203226 & 0.560 & 3.70 & 11990 & 256.79 (1,3), 298.06 (1,4) & 0.30 \\ 
0600589802 & 0.609 & 4.56 & 11560 & 952.54 (1,19), 866.63 (1,17), 560.32 (2,18), 969.17 (2,35), 479.04 (2,16) & 0.76 \\
0631161222 & 0.609 & 5.44 & 11430 & 707.54 (1,13), 678.27 (1,12), 466.56 91,7), 402.97 (1,6), & 0.60 \\
           &       &      &       &  368.25 91,5), 349.06 (2,10) & $\cdots$ \\ 
0631344957 & 0.593 & 5.14 & 11230 & 362.62 (1,5), 313.35 (1,4) & 0.53\\
0640201450 & 0.686 & 4.43 & 12280 & 254.88 (1,4) & $\cdots$\\
0661119673 & 0.570 & 4.55 & 11600 & 626.42 & $\cdots$\\
0762000503 & 0.646 & 4.12 & 11830 & 453.37 (1,8), 299.40 (1,5), 1429.62 (2,57) & 0.98 \\
0775564285 & 0.525 & 3.62 & 11600 & 852.63 (1,16), 628.41 (1,11) 895.85 (1,17) & 0.33 \\
0800126377 & 0.570 & 4.84 & 11550 & 325.04 (1,4) & $\cdots$\\
0800420812 & 0.593 & 4.84 & 11470 & 507.74 (1,8) & $\cdots$\\
0842451090 & 0.534 & 3.97 & 12000 & 311.72 (1,4) & $\cdots$\\ 
0900228144 & 0.686 & 4.87 & 11660 & 579.33 (1,11), 826.69 (1,17), 542.95 (2,19) & 0.59 \\
0900762564 & 0.525 & 3.62 & 11930 & 268.28 (1,2), 885.66 (1,17) & 0.16 \\
0902514572 & 0.548 & 3.68 & 11370 & 830.59 (1,16), 653.86 (1,12), 486.96 (1,8) & 0.41 \\
1102242692 & 0.609 & 5.44 & 11200 & 1009.13 (1,19) & $\cdots$\\
1102346472 & 0.548 & 7.33 & 11470 & 458.04 (1,6), 1489.63 (2,45) & 0.01 \\
1860439362 & 0.621 & 4.04 & 11760 & 574.97 (1,11), 462.26 (1,8) & 0.55 \\
1944049427 & 0.593 & 7.34 & 11370 & 606.26 (1,9) & $\cdots$\\
2045970633 & 0.690 & 5.23 & 11560 & 883.35 (1,18), 930.00 (1,19) & 0.26 \\
\hline
\end{tabular}
\label{tab:list1-res-1}
\end{table*}

\subsection{Rotation}

White dwarf stars are generally considered slow rotators, with rotation periods between a few hours and several days \citep[see for instance][]{2017EPJWC.15201011K}. From a sample of 116 white dwarfs, \citet{DaRosa24} found that the most probable rotation period is 3.9~h.
By considering the frequency splitting, we can estimate the rotation period of the white dwarf star, following equation \citep{1949ApJ...109..149C, 1951ApJ...114..373L}:
\begin{equation}
    \frac{1}{P_{\rm rot}} = \frac{\Delta \nu_{k,\ell, m}}{m (1-C_{k\ell})},
    \label{period}
\end{equation}
where $m$ is the azimuthal number, $\Delta \nu_{k,\ell, m}$ is the frequency separation and $C_{k\ell}$ is the rotational splitting coefficient given by:
\begin{equation}
    C_{k,\ell}=\frac{\int_0^{R_*} \rho r^2 [2\xi_r\xi_t + \xi_t^2] dr}{\int_0^{R_*} \rho r^2[\xi_r^2 + \ell(\ell + 1)\xi_t^2]dr},
\end{equation}
where $\rho$ is the density, $r$ is the radius, and $\xi_r$ and $\xi_t$ are the radial and horizontal displacement of the material. 

Considering that all our datasets have very high duty cycles, the TESS data are generally free of aliasing, and patterns of even frequency spacing in the Fourier transform can appear, which opens the opportunity to estimate the stellar rotation rate (e.g., \citealt{2004IAUS..215..561K}). Identifying rotationally split multiplets can also lead to the identification of the spherical degree $\ell$ and azimuthal order $m$ of the modes present in the period spectrum (e.g., \citealt{1991ApJ...378..326W,1994ApJ...430..839W}).

Using the combined data from the first five years of TESS, we identified multiplets, both triplets and doublets, in the period spectrum of 10 objects studied in this work. For four objects, we detected more than one multiplet. When doublets appear, we consider that the component with the largest amplitude corresponds to the $m=0$ component of the multiplet, while the second component corresponds to a $m=\pm 1$ component, unless otherwise stated \citep{1995ApJS...96..545B}. A list of the 9 objects and all detected multiplets is presented in Table~\ref{rotation}, together with the frequency separation, the frequency of the central component $m=0$, the value of the rotational splitting coefficient $C_{k\ell}$ obtained from the representative asteroseismological model listed in Table~\ref{tab:list1-res}, the corresponding rotation period in hours, calculated from Equation~\ref{period}, and a label indicating whether the multiplet is a doublet (D) or a triplet (T). All $m=0$ modes are fitted with a theoretical $\ell =1$ mode, except the mode with a frequency of 4998.00~$\mu$Hz corresponding to TIC~0055650407, and the mode with frequency 2035.58~$\mu$Hz corresponding to TIC~0149863849, that are better fitted by a $\ell =2$ mode.

\begin{table}
	\centering
	\caption{List of objects with detected rotational splitting. For each object, we list the TIC number (Column~1) the frequency separation $\Delta \nu$ (Column~2), the frequency of the $m=0$ central component (Column~3), the value of the $C_{k\ell}$ obtained from the asteroseismological representative model (Column~4), and the mean rotation period (Column~5). The last column indicates if the observed multiplet is a doublet (D) or a triplet (T).}
	\label{rotation}
	\begin{tabular}{cccccc} 
\hline
TIC     & $\Delta \nu$ [$\mu$Hz] & $\nu_{m=0}$ [$\mu$Hz] & $C_{k\ell}$ & $\Bar{P}_{\rm rot}$ [h] & \\
\hline
0007675859 & 22.66 & 2830.83 & 0.4172 & 7.15  & D\\ 
0055650407 & 11.79 & 7883.95 & 0.4853 & 12.12  & D\\
 $\cdots$  & 14.24 & 4998.00 & 0.1012 & 17.54  & T\\
 $\cdots$  & 12.38 & 3810.10 & 0.4717 & 11.85  & D\\
 $\cdots$  & 13.48 & 3131.08 & 0.3817 & 12.74  & D\\
0079353860 & 8.51 & 1902.77 & 0.4775 & 17.06 & D \\ 
0149863849 & 16.15 & 2035.58 & 0.1574 & 14.49 & D \\
$\cdots$   & 6.25 & 1316.17 & 0.4907 & 22.64 & T \\
$\cdots$   & 6.22 & 1250.00 & 0.4910 & 22.73 & T \\
0201860926 & 16.51 & 1135.41 & 0.4865 & 8.64 & D \\
$\cdots$   & 8.48 & 1912.56 & 0.4938 & 16.58 & T \\
0343296348 & 12.20 & 3468.97 & 0.4576 & 12.35  & T\\
0353727306 & 32.64 & 2470.11 & 0.4871 & 4.65 & D \\
$\cdots$   & 30.74 & 2126.57 & 0.4853 & 4.37 & D \\
0800126377 & 11.94 & 3076.46& 0.3512 & 15.10   & D\\
1102242692 & 5.78 & 991.05 & 0.4900 & 24.14 & T \\
\hline
\end{tabular} 
\end{table}

\begin{figure}
  \includegraphics[width=0.45\textwidth]{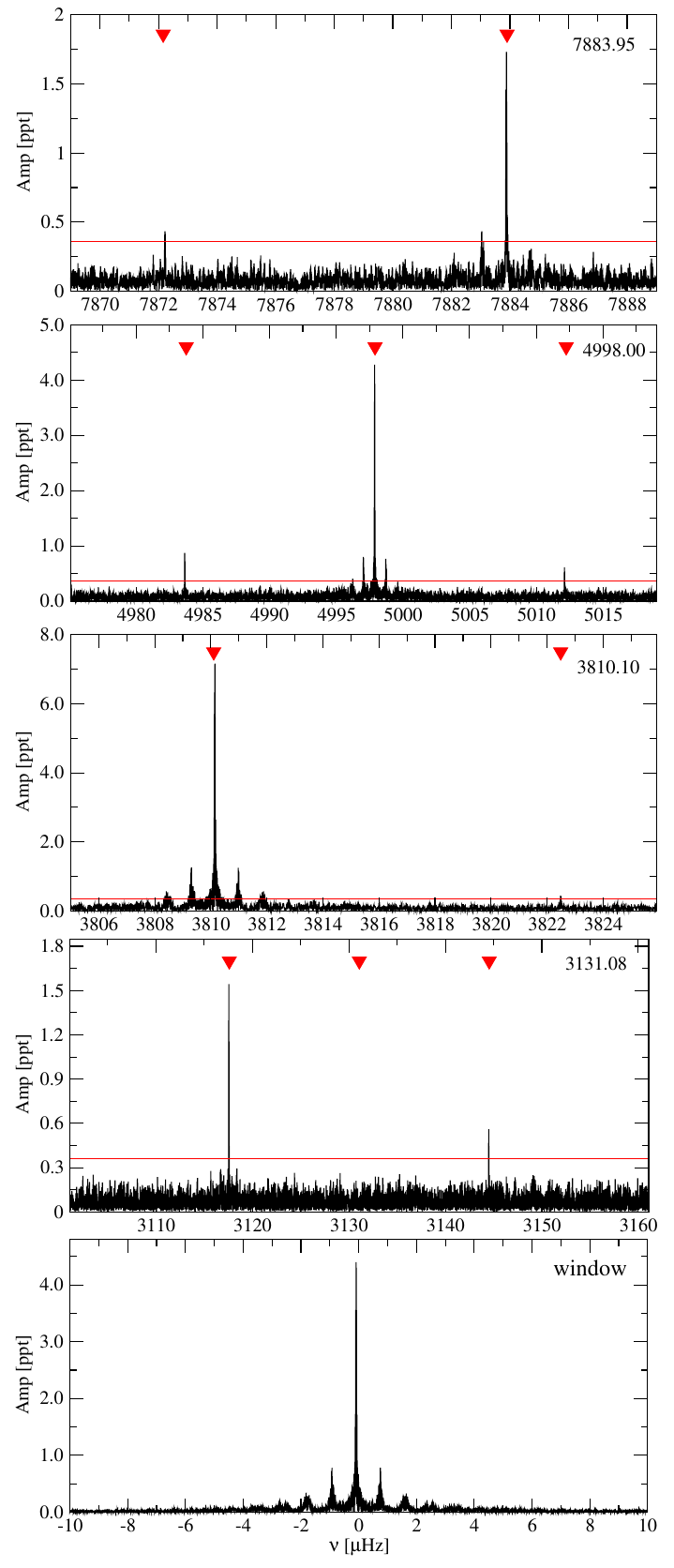}
    \caption{Detailed look at the FT for TIC~0055650407 around the detected multiplets centered around 7883.95, 4998,00, 3131,08 and 3117.60 $\mu$Hz from top to bottom (see Table~\ref{rotation}). The triangles indicate the position of the components, while the horizontal line indicates the amplitude detection limit. The bottom panel shows the corresponding spectral window for the data, with $\simeq 1\mu$Hz aliases. } 
    \label{rotation-example}
\end{figure}

 Figure~\ref{rotation-example} shows the FT for the four multiplets detected for TIC~0055650407. For this object, we consider the components of the doublet at $\sim 262$~s to correspond to the components of $m = \pm 1$, and thus the multiplet would be centered at a frequency of 3131.08~$\mu$Hz. The period with $\sim$200~s (4998.00~$\mu$Hz) is identified as a $\ell=2$ mode in the representative asteroseismological model, giving a rotation period of 17.54~h. If we consider an alternative identification, as an $\ell$=1, k=2 mode, with a $C_{k\ell}$ = 0.3719, the resulting rotation period would be 12.25~h, closer to the values obtained for the other three multiplets observed for this object.

TIC~0149863849 has a complex structure of multiplets centered at 759.78~s and 800.00~s. In addition, there are two doublets, the more prominent being the one corresponding to the largest amplitude peak at 398.96~s. 

TIC~0201860926 shows a doublet and a triplet that lead to rotation periods of 8.64 and 16.58~h. Note that if we identify the two components of the doublet as $m=\pm 1$ both multiplets would indicate a similar rotation period. 

A second possible solution is listed for TIC~0353727306 in Table~\ref{tab:list1-res} where the periods identified as multiplets are represented by theoretical modes with $\ell$ = 2. If we consider this solution, the mean rotation period for this object is 7.45~h (14.90~h), when the components are considered to be $m=1$ ($m=2$). 

The FT showing in detail de frequency range for all detected multiplets listed in Table \ref{rotation} can be found in Appendix \ref{apendix}.

\begin{figure}
	\includegraphics[width=0.6\textwidth]{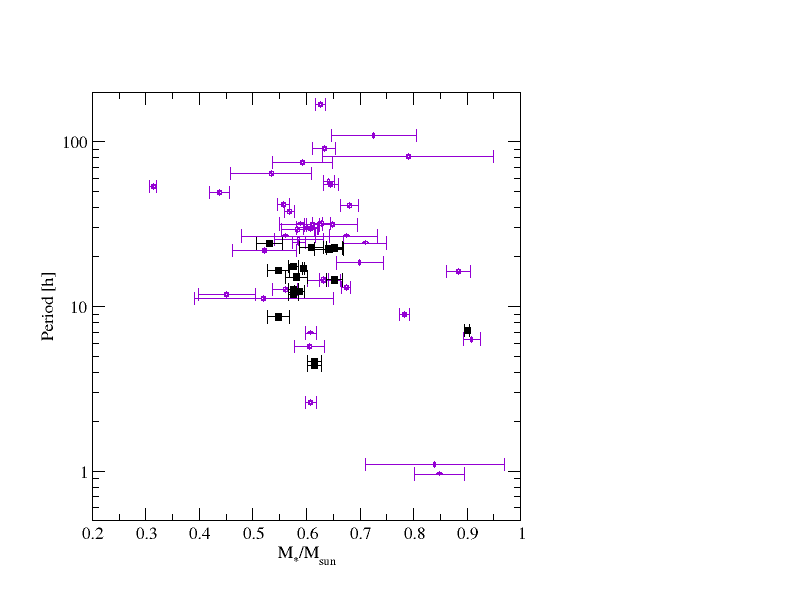}
    \caption{Rotation period obtained from asteroseismology as a function of stellar mass for DA white dwarfs (see Table~\ref{tab:list1}). Purple dots correspond to values from the literature \citep{Romero22, 10.1093/mnras/sts438, Bell_2017, 10.1093/mnras/stt2069, Hermes17, Fu2019,Giammichele_2016, 10.1093/mnras/stt2420,  Kawaler15, Castanheira2013,refId0, Pech2006,Bradley_2001, teste, 1995ApJ...447..874K, Li2017}, while black squares correspond to the data from Table~\ref{rotation}. For the objects with more than one multiplet we include all rotation period determinations.} 
    \label{rotation-plot}
\end{figure}

Figure~\ref{rotation-plot} shows the rotation period distribution obtained from asteroseismology as a function of stellar mass for the sample of DA white dwarf stars. The purple dots correspond to data from the literature, while the black squares correspond to the values listed in Table~\ref{rotation}. For objects with more than one multiplet, we plot all values. Note that the values for the rotation periods agree well with those in the literature.

\section{Analysis of the sample }
\label{section6}

In this section, we analyze the results from the periods and structure parameter for the 103
pulsating DA white dwarf stars that have been identified from the TESS data in the first five years.  

\subsection{Stellar mass and effective temperature}

\begin{figure}
	\includegraphics[width=0.6\textwidth]{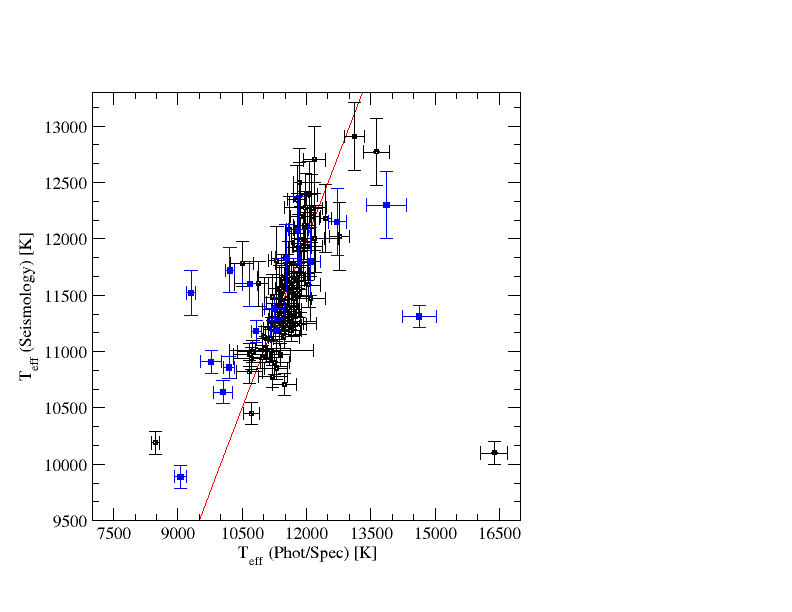}
    \caption{Comparison between the photometric or spectroscopic  effective temperature (Table~\ref{tab:list1}) and the asteroseismological effective temperature (Table~\ref{tab:list1-res}). The red line indicates the 1:1 correspondence lines. Blue squares (black circles) correspond to objects with photometric or spectroscopic masses below (above) 0.49~$M_{\odot}$. } 
    \label{teff-comp}
\end{figure}

Figure~\ref{teff-comp} shows the comparison for the effective temperature values determined from Gaia photomety + parallax or spectroscopy (x axis) and asteroseismology (y axis). The uncertainties for the photometric or spectroscopic determinations are taken from Table~\ref{tab:list1}. The internal uncertainties for the asteroseismological values are fixed by the resolution of the model grid to be 100~K, 200~K, and 300~K, for temperatures below $11\,400$ K, between $11\,400$ K and $11\, 800$ K and above $11\,800$~ K, respectively. The 16 objects with photometric or spectroscopic stellar masses below 0.49~$M_{\odot}$, which is the minimum mass of our C/O core model grid, are depicted as blue squares, while those with stellar masses above this value are depicted with black circles. As can be seen in Figure~\ref{teff-comp}, the data cluster around the 1:1 correspondence line in red. Most of the outliers correspond to candidates for low-mass white dwarfs, with photometric effective temperatures above or below the blue and red edges of the instability strip, respectively \citep{2015ApJ...809..148T,Hermes17}. 

TIC~0415337224 shows a photometric effective temperature of $\sim$$16{,}400$\,K, but the periods are long, which are characteristic of red edge pulsators \citep{2006ApJ...640..956M}. TIC~0103700861 shows the lowest effective photometric temperature of the sample, $\sim$$8500$~K, with two periods around 500 and 700 s, characteristic of warm--like to coll pulsators, in agreement with the asteroseismological determination of the effective temperature. Both stars may have contaminated Gaia photometry.

\begin{figure}
 	\includegraphics[width=0.6\textwidth]{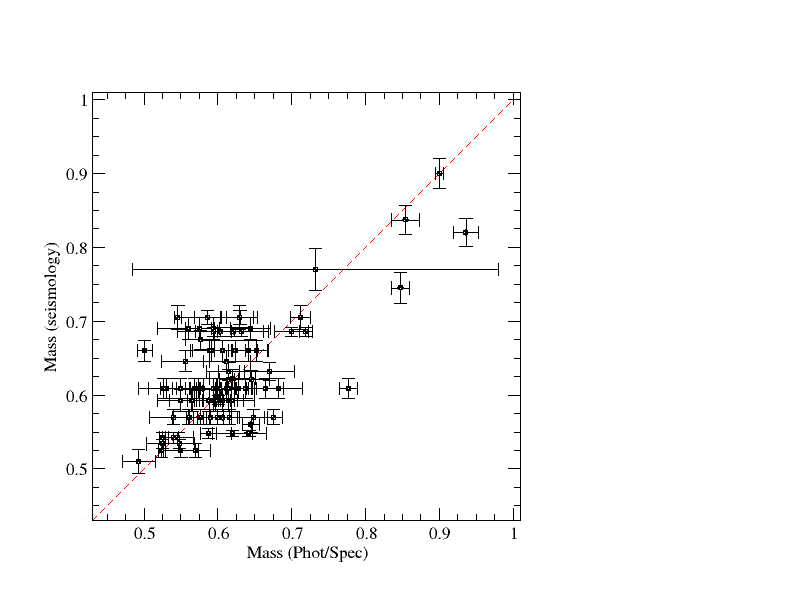}
    \caption{Comparison between the stellar mass obtained from photometry or spectroscopy (Table~\ref{tab:list1}) and the asteroseismological fit (see Table~\ref{tab:list1-res}). The red line indicates the 1:1 correspondence. Objects with photometric or spectroscopic masses below 0.49\,$M_{\odot}$ are not included. } 
    \label{mass-comp}
\end{figure}

In Figure~\ref{mass-comp} we compare the value of the stellar mass from photometry + Gaia parallax or spectroscopy (x axis) and from asteroseismology (y axis). We do not include objects with stellar masses below 0.49\,$M_{\odot}$, which is the minimum mass of our C/O core model grid. Although the points are around the 1:1 correspondence line, there is a large scatter. The spread around the 1:1 correspondence line for stellar masses between 0.5 and 0.7\,$M_{\odot}$ can be explained by means of the hydrogen envelope mass. For the same stellar mass, the radius of the star is smaller the less hydrogen is left in the outer layers \citep{2019MNRAS.484.2711R}, causing a degeneracy in the mass-radius relation. If only a canonical thick hydrogen envelope mass is considered to estimate the stellar mass from photometry and parallax, a thin envelope white dwarf will appear as a smaller and thus more massive star. The median of the stellar mass of the sample considered in Figure \ref{mass-comp} is $\left<M_{\rm ph/sp}\right> =$ 0.602\,M$_{\odot}$ for the values obtained by photometry or spectroscopy, while for the seismological mass, we obtain a value of $\left<M_{\rm seis}\right> =$ 0.609 M$_{\odot}$.

\subsection{Weighed mean period }

From the observed period spectrum of the 103 variable DAV stars, we can compute the weighted mean period (WMP) as:

\begin{equation}
    {\rm WMP}=\frac{\sum_i P_i A_i}{\sum_i A_i},
    \label{WMP-eq}
\end{equation}
where $P_1$ is the observed period and $A_i$ its amplitude. \citet{2006ApJ...640..956M}, using the available sample of known ZZ Cetis at the time and atmosphere parameters obtained from spectroscopy, found a relation between the WMP and the effective temperature. According to their results, the value of the WMP increases for lower effective temperatures. In the following, we revisit this relation using the sample of 103 DAVs discovered from TESS data. 

\begin{figure}
	\includegraphics[width=0.5\textwidth]{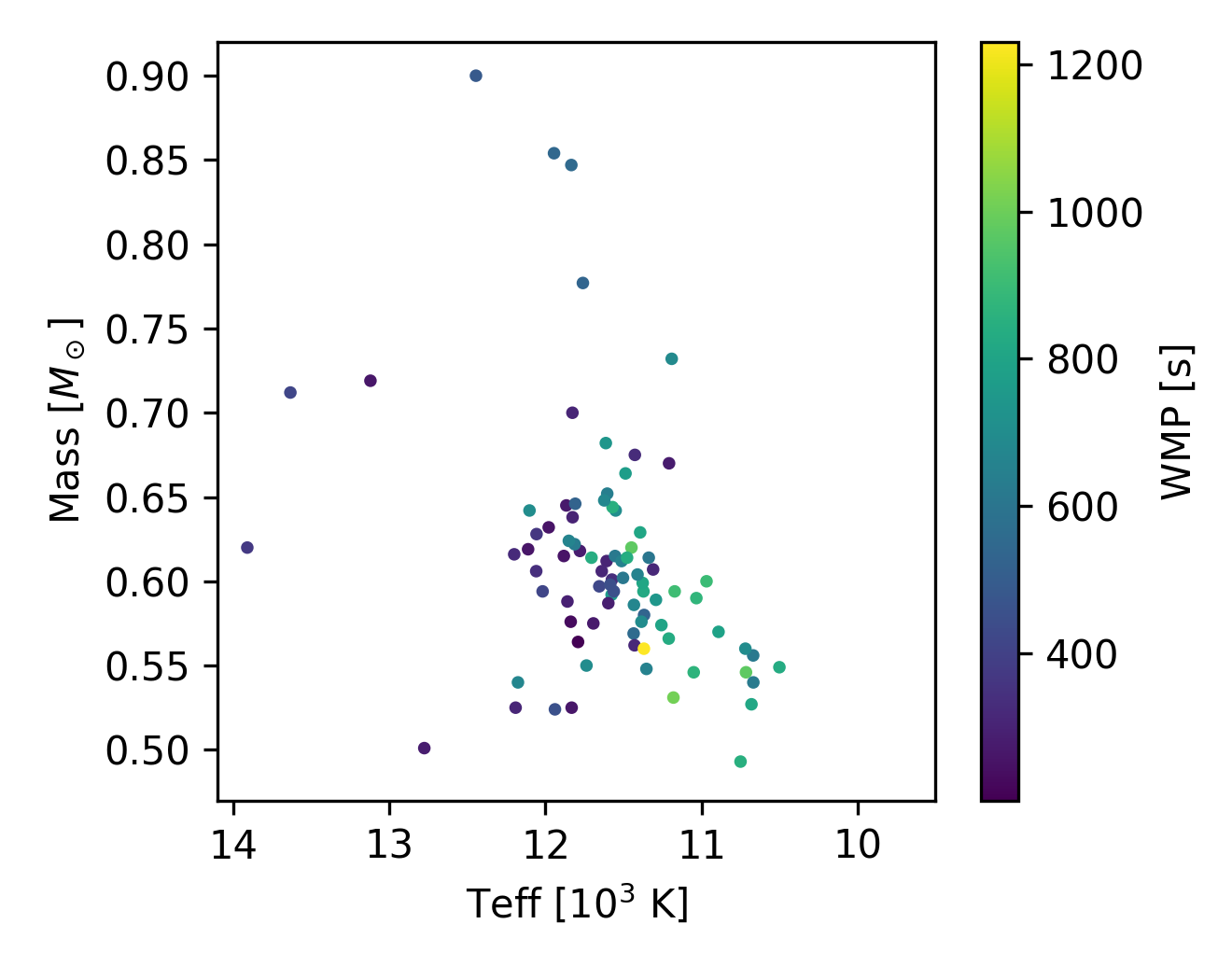}
 	\includegraphics[width=0.5\textwidth]{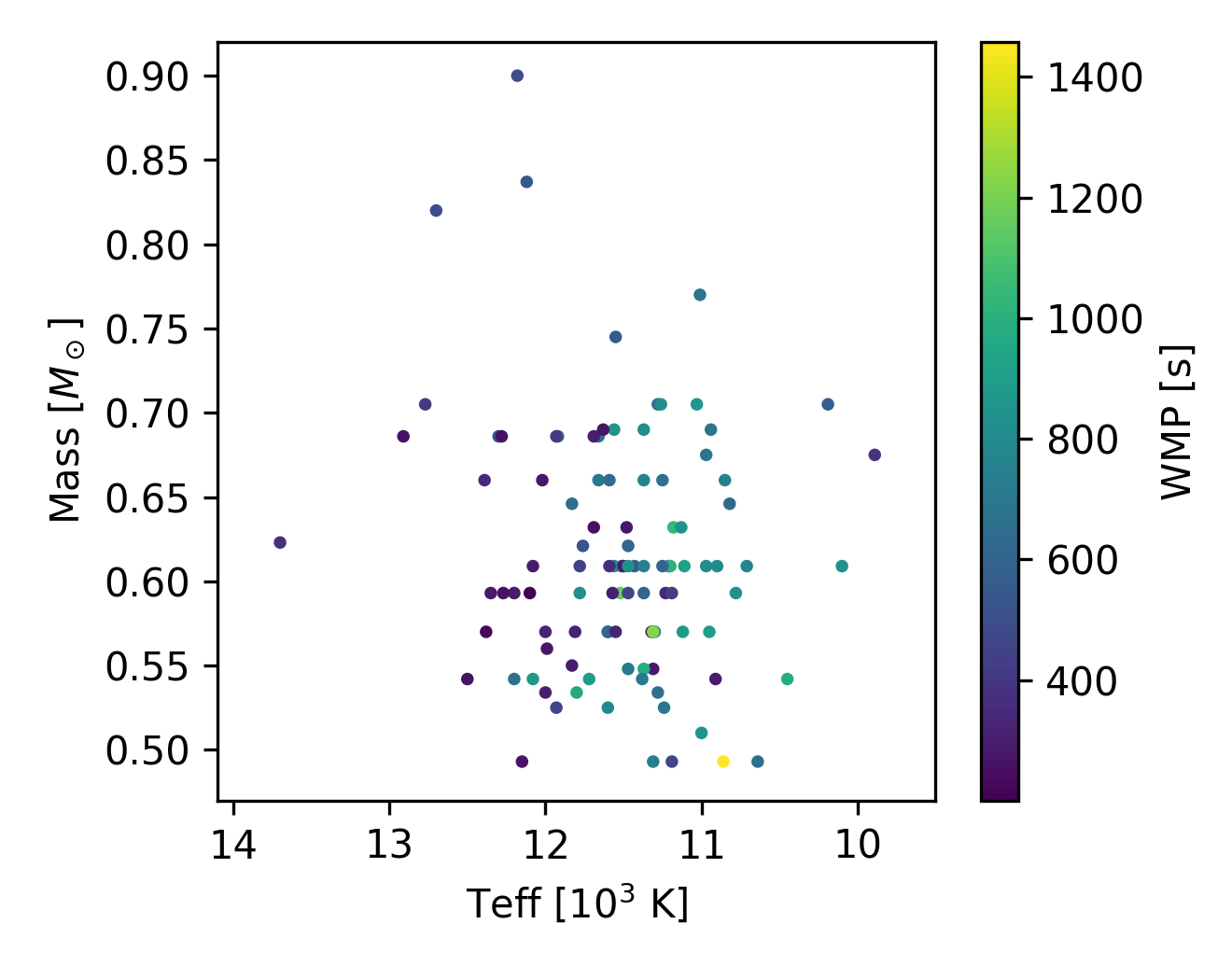}
    \caption{Value of the weighted mean period (WMP) as a function of photometric or spectroscopic (top panel) and seismological (bottom panel) stellar mass and effective temperature. The value of the WMP is shown in the color scale.} 
    \label{WMP}
\end{figure}

The WMP as a function of stellar mass and effective temperature is depicted in Figure~\ref{WMP}, for values obtained from photometry or spectroscopy (top panel) and the asteroseismological fit (bottom panel). The value of the WMP is on a color scale. We consider objects with photometric or spectroscopic stellar masses larger than 0.49~M$_{\odot}$ and effective temperatures between $10{,}000$ and $14{,}000$ K. From these figures, we found a dependence of the WMP on the effective temperature, which is shorter for higher temperatures, in agreement with the results presented in \citet{2006ApJ...640..956M}. The value of the Pearson coefficient is $-$0.51 for the photometric or spectroscopic effective temperature determinations, indicating a moderate correlation of the WMP with this parameter. In the case of the effective temperature obtained from the asteroseimological fits, the Pearson coefficient is $-$0.57, also indicating a moderate correlation. Assuming a linear dependence of the WMP with the effective temperature, we obtain the following relations.

\begin{equation}
    {\rm WMP} (s) = -0.197\times T_{\rm eff}(K) + 2876
\end{equation}

\noindent for photometric or spectroscopic effective temperatures, and

\begin{equation}
    {\rm WMP} (s) = -0.267\times T_{\rm eff}(K) + 3680
\end{equation}

\noindent for effective temperature obtained from asteroseismology, with $T_{\rm eff}$ in $K$ and the WMP in seconds. Finally, there is no clear dependence of the WMP on stellar mass.

\begin{figure}
	\includegraphics[width=0.6\textwidth]{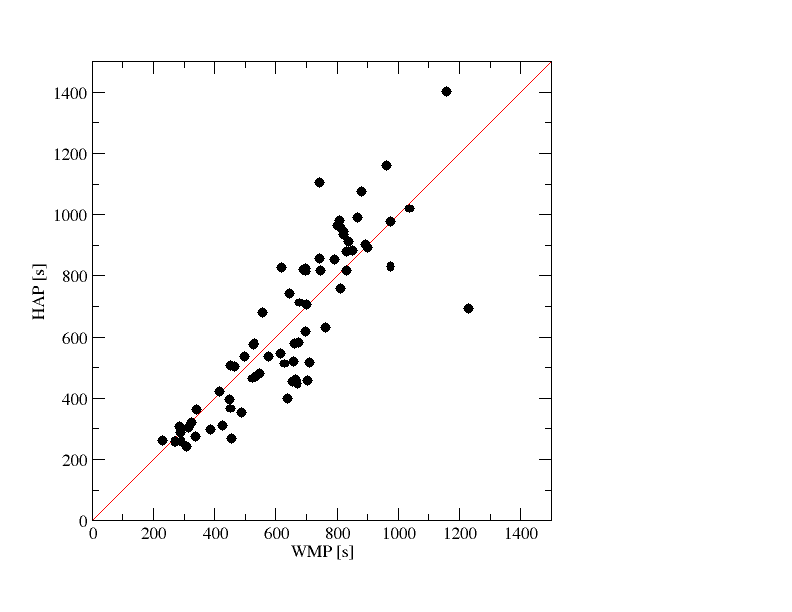}
     \caption{Comparison between the values of the weighed mean period (WMP) and the highest amplitude period (HAP). Objects with only one period are not included. The red line indicate the 1:1 relation. } 
    \label{WMP-HAP-comp}
\end{figure}

The WMP is usually dominated by the period with the highest amplitude, especially when the amplitude differences are large. However, if the period amplitudes are similar, the WMP can differ from the highest-amplitude period (HAP). Figure~\ref{WMP-HAP-comp} shows the comparison between the WMP (x-axis) and the HAP (y-axis), for objects with more than one observed period. As expected, the data values are close to the 1:1 correspondence line in red, but a dispersion is observed from the plot. For periods around $\sim$700 s, the WMP value tends to be higher than the HAP value, while for periods longer than $\sim$700 s the WMP is shorter than the HAP. The Pearson coefficient is 0.85, indicating a strong correlation. 

\begin{figure}
	\includegraphics[width=0.5\textwidth]{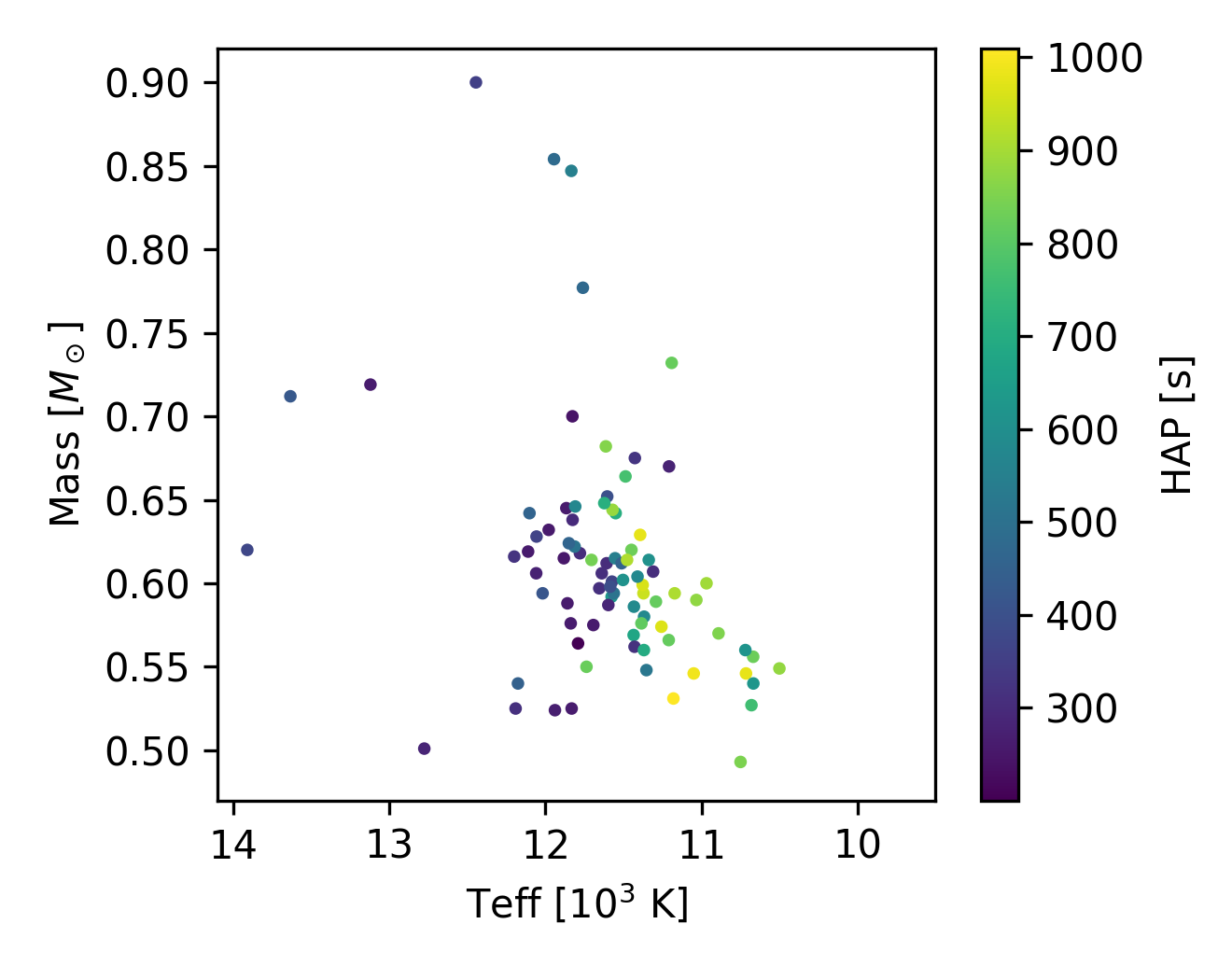}
 	\includegraphics[width=0.5\textwidth]{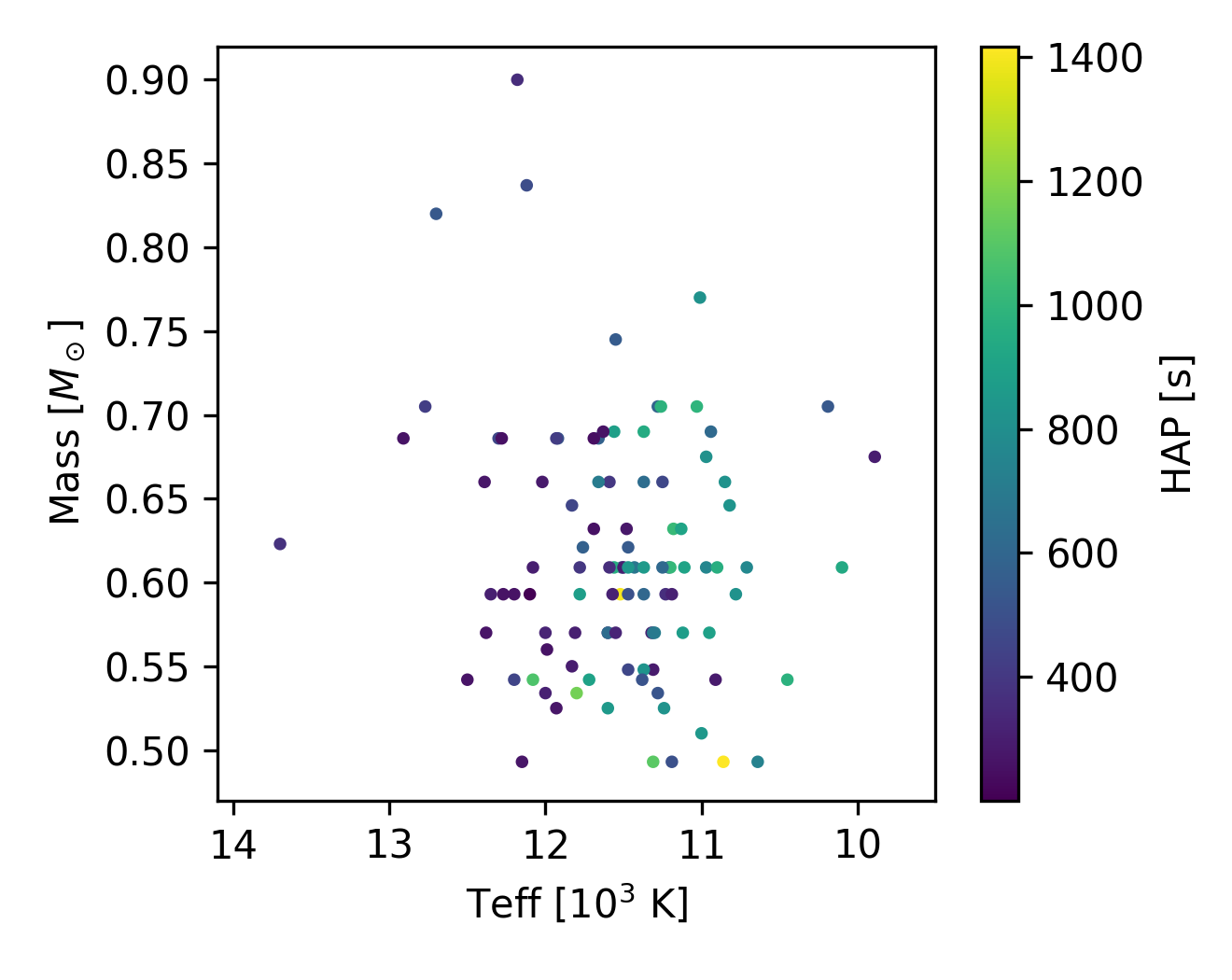}
    \caption{Value of the highest amplitude period (HAP) as a function of photometric or spectroscopic (top panel) and seismological (bottom panel) stellar mass and effective temperature. The value of the HAP is shown in the color scale.} 
    \label{HAP}
\end{figure}

In Figure~\ref{HAP} we depict the value of the HAP as a function of effective temperature and stellar mass obtained from photometry or spectroscopy and asteroseismology in the top and bottom panels, respectively. The value of the HAP is represented on the color scale. In the figure, we note that the value of the HAP also increases with decreasing effective temperatures, following the same trend as the WMP. The Pearson coefficient is $-$0.52 and $-$0.57, when we consider the effective temperatures from photometry or spectroscopy and asteroseismology, respectively, indicating a moderate correlation of the HAP with this parameter. 

There is also a dependence with stellar mass for effective temperatures higher than $\sim$$12{,}000$~ K, where the HAP values increase with stellar mass. However, given that there are few objects with stellar masses larger than $\sim$0.7$\,M_{\odot}$, this dependence is not well grounded. 

\section{Conclusions}
\label{conclusions}
In the first five years, TESS has already proven to be a powerful tool in discovering new pulsating white dwarfs. Although it is a small telescope with low spatial resolution, its excellent time span has allowed not only the discovery of more than 100 new pulsating white dwarfs but also the detection of multiplets in many of them.

In this work, we present the discovery of 32 new pulsating DA white dwarf stars, based on observations of the TESS mission from Sector 40 to Sector 69. In addition, we revisited 21 ZZ Cetis reported by \citet{Romero22} that present differences in their period spectra, after observations from Cycles 4 and 5 were considered. The complete sample of pulsating DA white dwarfs discovered using TESS observations is now 103 objects.

For each object, we performed an asteroseismological study and determined the main structure parameters such as stellar mass, effective temperature, and hydrogen envelope mass. In cases where only one or two periods were detected, we use the photometric or spectroscopic observations as an additional constraint to the fit. 

We detected a component of rotational splitting for 9 objects, with rotation periods from 4~h to 1~d, in line with the observed rotation period distribution obtained from asteroseismology for other known DA white dwarf stars. For four objects, we detected more than one multiplet that could open the opportunity to study radial differential rotation. The median for the rotation period for the 9 objects reported in this work is 13.6~h.

The median value for the stellar mass for the sample of ZZ Ceti from TESS from photometry and/or spectroscopy and seismology is $\left<M_{\rm ph/sp}\right> = $ 0.602 M$_{\odot}$ and $\left<M_{\rm seis}\right> = 0.609$  M$_{\odot}$, respectively. These values are in line with the mean stellar mass of 351 known ZZ Ceti stars of $\left<M_{\rm seis}\right> = 0.644 \pm 0.034 M_{\odot}$ \citep{Romero22}.

Finally, we revisit the relation between the weighted mean period and the effective temperature. We found a moderate correlation for both the photometric or spectroscopic and the seismological effective temperatures with the WMP value. We did not find a strong dependence on the stellar mass.

 

%




\section{Appendix}
\label{apendix}

In this appendix we present the Fourier Transform around the frequency range where the multiplets are detected for all objects listed in Table \ref{rotation}, except for TIC~0055650407. The horizontal red line represents the amplitude detection limit. All plots are labeled by the central frequency of the multiplet. 

\begin{figure}
	\includegraphics[width=0.55\textwidth]{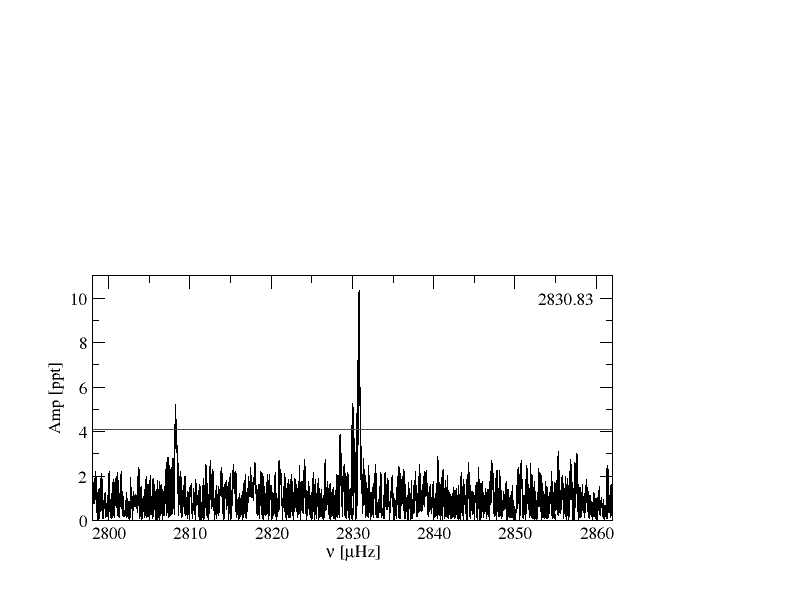}
 	    \caption{FT showing the doublet for TIC~0007675859} 
    \label{ap-1}
\end{figure}

\begin{figure}
	\includegraphics[width=0.55\textwidth]{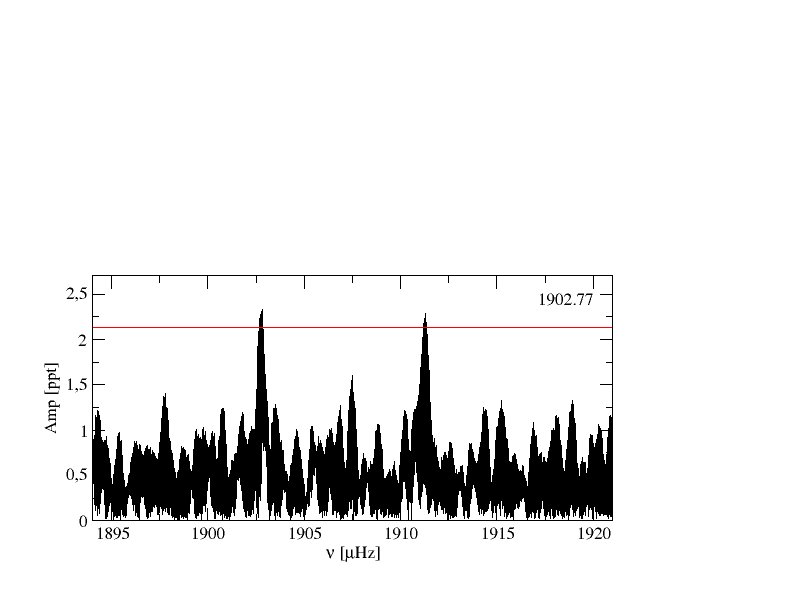}
 	    \caption{FT showing the doublet for TIC~0079353860.} 
    \label{ap-2}
\end{figure}

\begin{figure}
	\includegraphics[width=0.55\textwidth]{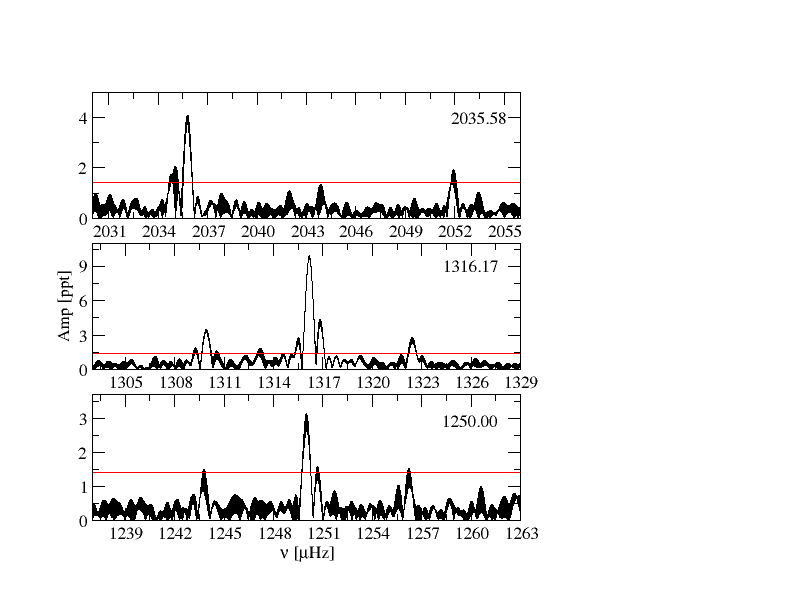}
 	    \caption{FT showing the multiplets for TIC~0149863849. } 
    \label{ap-3}
\end{figure}

\begin{figure}
	\includegraphics[width=0.55\textwidth]{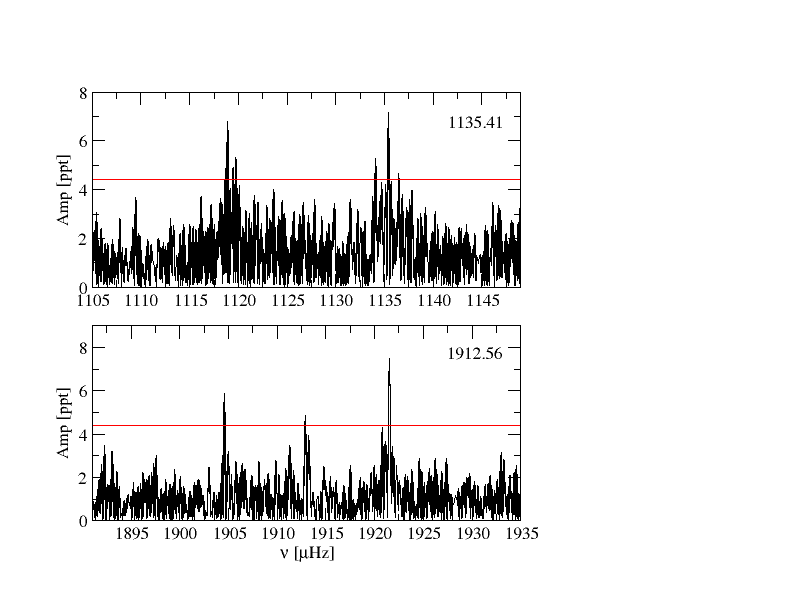}
 	    \caption{FT showing the multiplets for TIC~0201860926. } 
    \label{ap-4}
\end{figure}

\begin{figure}
	\includegraphics[width=0.55\textwidth]{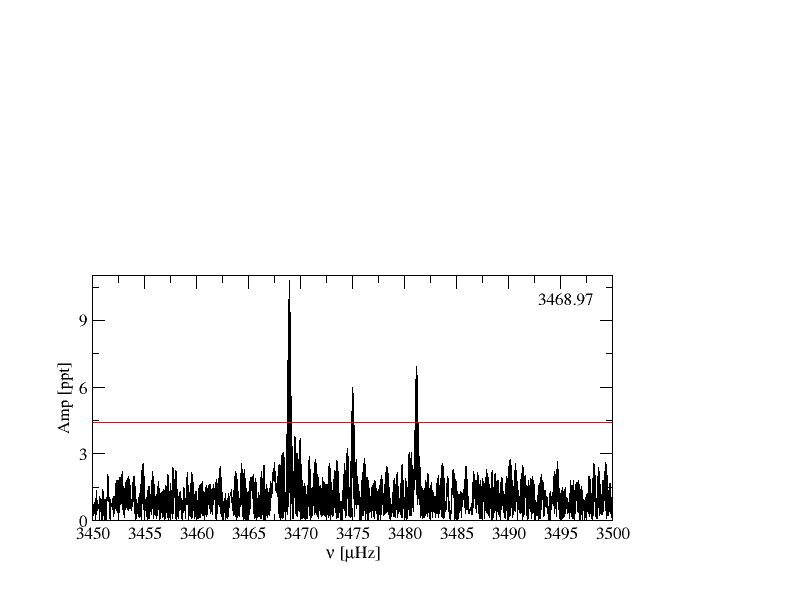}
 	    \caption{FT showing the triplet for TIC~0343296348. } 
    \label{ap-5}
\end{figure}

\begin{figure}
	\includegraphics[width=0.55\textwidth]{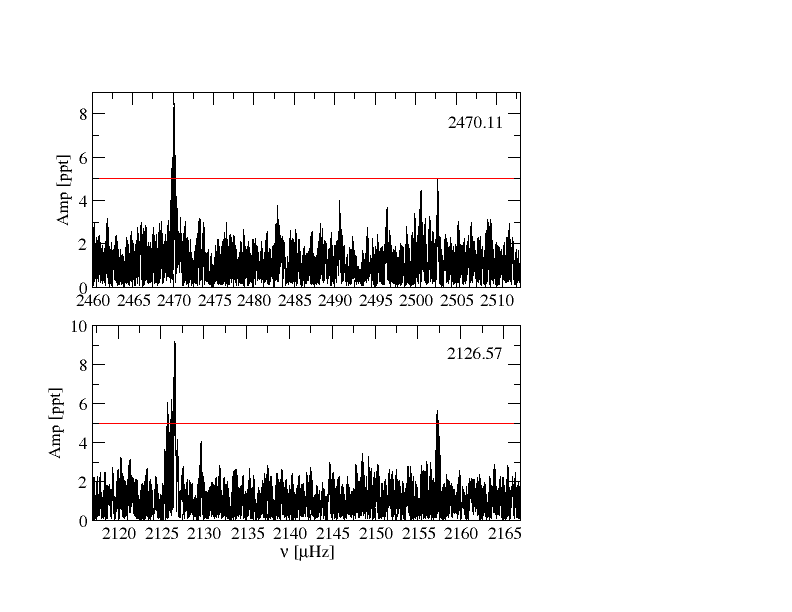}
 	    \caption{FT showing the doublets for TIC~0353727306. } 
    \label{ap-6}
\end{figure}

\begin{figure}
	\includegraphics[width=0.55\textwidth]{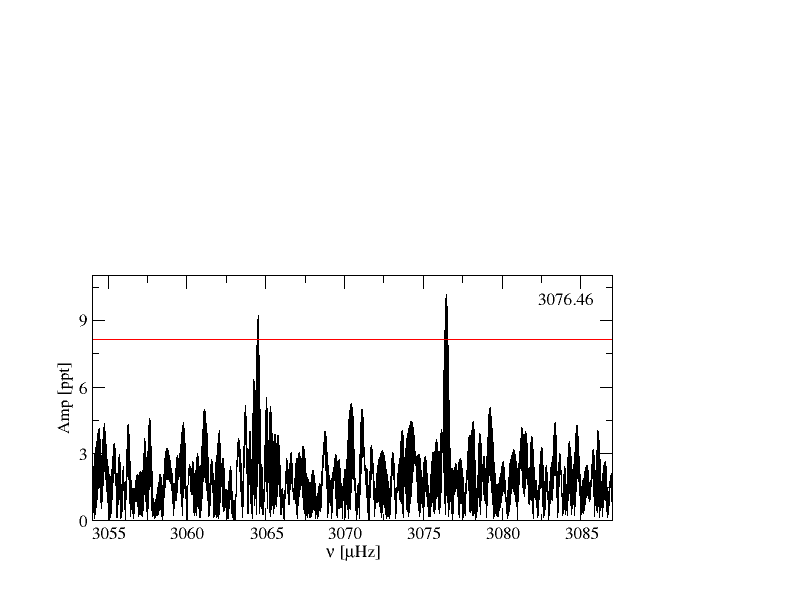}
 	    \caption{FT showing the doublet for TIC~0800126377. } 
    \label{ap-7}
\end{figure}

\begin{figure}
	\includegraphics[width=0.55\textwidth]{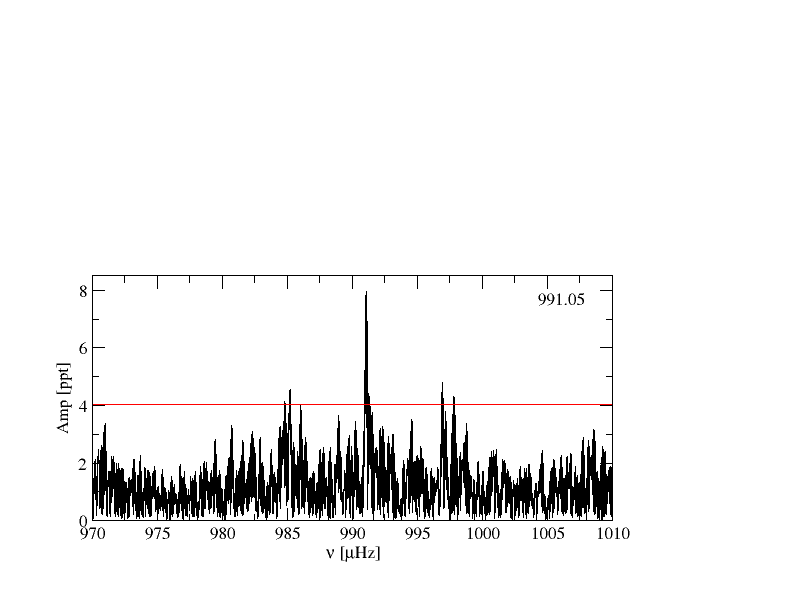}
 	    \caption{FT showing the triplet for TIC~1102242692. } 
    \label{ap-8}
\end{figure}

\bibliography{NewTESS}{}
\bibliographystyle{aasjournal}



\end{document}